\documentclass[a4paper,11pt]{article}
\pdfoutput=1 

\usepackage{jcappub,wrapfig,verbatim,amsthm,graphicx,hyperref,subcaption,caption} 
                     
\usepackage[utf8]{inputenc}

\newcommand{\f}{f_{\mu\nu}}
\newcommand{\g}{g_{\mu\nu}}
\newcommand{\Vg}{V^\mu{}_\nu}
\newcommand{\Vf}{\widetilde{V}^\mu{}_\nu}
\newcommand{\Gg}{G^\mu{}_\nu}
\newcommand{\Gf}{\widetilde{G}^\mu{}_\nu}
\newcommand{\Tg}{T^\mu{}_\nu}

\newcommand{\wt}{\widetilde}

\newcommand{\mfphat}{m_\mathrm{FP}}
\newcommand{\omegaleff}{\Omega_\Lambda}

\newcommand{\omegade}{\Omega_\mathrm{DE}}
\newcommand{\omegadef}{\wt{\Omega}_\mathrm{DE}}

\newcommand{\mfp}{m_\mathrm{FP}}

\newcommand{\omegal}{\Omega_\Lambda}
\newcommand{\omegamnot}{\Omega_{m,0}}

\title{\boldmath Observational constraints on bimetric gravity}

\author{Marcus Högås, Edvard Mörtsell}

\affiliation{The Oskar Klein Centre, Department of Physics, Stockholm University, SE 106 91, Stockholm, Sweden}

\emailAdd{marcus.hogas@fysik.su.se, edvard@fysik.su.se}

\abstract{Ghost-free bimetric gravity is a theory of two interacting spin-2 fields, one massless and one massive, in addition to the standard matter particles and fields, thereby generalizing Einstein's theory of general relativity. To parameterize the theory, we use five observables with specific physical interpretations. We present, for the first time, observational constraints on these parameters that: (i) apply to the full theory, (ii) are consistent with a working screening mechanism (i.e., restoring general relativity locally), (iii) exhibit a continuous, real-valued background cosmology (without the Higuchi ghost). For the cosmological constraints, we use data sets from the cosmic microwave background, baryon acoustic oscillations, and type Ia supernovae. Bimetric cosmology provides a good fit to data even for large values of the mixing angle between the massless and massive gravitons. Interestingly, the best-fit model is a self-accelerating solution where the accelerated expansion is due to the dynamical massive spin-2 field, without a cosmological constant. Due to the screening mechanism, the models are consistent with local tests of gravity such as solar system tests and gravitational lensing by galaxies. We also comment on the possibility of alleviating the Hubble tension with this theory.}

\begin{document}
\maketitle
\flushbottom

\section{Introduction}
Ghost-free bimetric gravity is a theory of two interacting spin-2 field, one massive and one massless, in addition to the standard matter particles and fields. To achieve this setup, one must introduce a second symmetric spin-2 field (i.e., a metric) $\f$ besides the physical metric $\g$. The physical metric is the one to which standard matter particles and fields couple.

Among the virtues of bimetric gravity are self-accelerating cosmological models, that is without a cosmological constant \cite{Volkov:2011an,vonStrauss:2011mq,Comelli:2011zm,Volkov:2012wp,Volkov:2012zb,Akrami:2012vf,Volkov:2013roa,Konnig:2013gxa}. For these models, the accelerated expansion of the Universe is due to the interaction between the spin-2 fields which contributes an extra term, $\omegade$, in the cosmological equations of motion. Interestingly, for these models, a small value of $\omegade$ is technically natural in the sense of 't Hooft, which means that its value is protected from quantum corrections \cite{tHooft:1979rat}. Bimetric cosmology can push the Hubble constant inferred from cosmic microwave background observations in the right direction in order to ease the tension with local measurements \cite{Mortsell:2018mfj}. The theory has a screening mechanism that can restore general relativity on solar system scales \cite{Babichev:2013pfa,Enander:2015kda,Platscher:2018voh}. Among the challenges is the existence of a gradient instability for linear perturbations around cosmological background solutions \cite{Comelli:2012db,Khosravi:2012rk,Berg:2012kn,Sakakihara:2012iq,Konnig:2014dna,Comelli:2014bqa,DeFelice:2014nja,Solomon:2014dua,Konnig:2014xva,Lagos:2014lca,Konnig:2015lfa,Aoki:2015xqa,Mortsell:2015exa,Akrami:2015qga,Hogas:2019ywm,Luben:2019yyx} and finding a stable (well-posed) form of the equations of motion in order to obtain long-term numerical evolution of generic systems \cite{Kocic:2018ddp,Kocic:2018yvr,Kocic:2019zdy,Torsello:2019tgc,Kocic:2019gxl,Torsello:2019jdg,Torsello:2019wyp,Kocic:2020pnm}.

Usually, bimetric theory is parameterized with five parameters, $\beta_n$ with $n=0,1,...,4$ (to be defined below). However, the $\beta$-parameters are not observables and the reported constraints on them depend on which convention is used, being different in different papers. Following Ref.~\cite{PhysParamTh}, which generalizes the works of Refs.~\cite{Luben:2019yyx,Luben:2020xll}, we use a parameterization in terms of five observables: the mixing angle $\theta$ between the massless and massive gravitons, the graviton (Fierz--Pauli) mass $\mfphat$, the effective cosmological constant $\omegaleff$, and two parameters, $\alpha$ and $\beta$ (not to be confused with the $\beta$-parameters $\beta_n$), that appear in the screening mechanism which cancels the extra gravitational forces on local scales (e.g., in the solar system).  

Earlier results have shown the theory to be compatible with cosmological data from the cosmic microwave background (CMB), baryon acoustic oscillations (BAO), and supernovae type Ia (SNIa) \cite{vonStrauss:2011mq,Akrami:2012vf,Konnig:2013gxa,Dhawan:2017leu,Lindner:2020eez,Luben:2020xll}, gravitational wave observations \cite{DeFelice:2013nba,Fasiello:2015csa,Cusin:2015pya,Max:2017flc}, solar system tests \cite{Platscher:2018voh,Luben:2018ekw}, velocity dispersion and strong gravitational lensing in galaxies \cite{Sjors:2011iv,Enander:2013kza,Enander:2015kda,Platscher:2018voh}, as well as gravitational lensing by galaxy clusters \cite{Platscher:2018voh}. Typically, in these works, only a subset of observations has been used or the results apply only to a restricted set of models, that is, setting one or several of the $\beta$-parameters to zero. Also, the chosen set of parameters are typically not compatible with a working screening mechanism or a continuous, real-valued cosmology (without the Higuchi ghost). As shown in Ref.~\cite{PhysParamTh}, these requirements can be formulated as a set of analytical constraints on the physical parameters. In this paper, for the first time, we derive the observational constraints on the full theory while at the same time guaranteeing that the above analytical constraints are satisfied.

\paragraph{Notation.} For the most part, we use geometrized units where Newton's gravitational constant and the speed of light are set to one, $G=c=1$. Quantities constructed from the second metric $\f$ are denoted with tildes, otherwise constructed from the physical metric $\g$. We use redshift $z$ as a time variable, defined via the equation $1+z=a_0 /a$ where $a_0$ is the scale factor today.

\section{Bimetric gravity}
\label{sec:BR}
The equations of motion are constructed to avoid the Boulware--Deser ghost which plagues general theories of massive gravity \cite{Boulware:1973my,Hassan:2011zd}. Assuming that there is only one matter sector, coupled to $\g$, the equations read,
\begin{equation}
	\label{eq:EoM}
	\Gg = \kappa_g (\Tg + \Vg), \quad	\Gf = \kappa_f \Vf,
\end{equation}
where $\Gg$ and $\Gf$ are the Einstein tensors, $\Tg$ is the standard matter stress--energy, and $\kappa_g$ and $\kappa_f$ are the gravitational constants of $\g$ and $\f$, respectively. We denote the ratio of the gravitational constants,
\begin{equation}
\kappa \equiv \kappa_g / \kappa_f.
\end{equation}
The bimetric stress--energies $\Vg$ and $\Vf$ contain the metrics $\g$ and $\f$ as well as five constant parameters $\beta_0,...\beta_4$ with dimension of curvature (see e.g. \cite{PhysParamTh} for the mathematical form of $\Vg$ and $\Vf$). The bimetric stress--energies are conserved, $\nabla_\mu \Vg = \wt{\nabla}_\mu \Vf = 0$, which follows from conservation of matter stress--energy, $\nabla_\mu \Tg=0$, and the Bianchi identities, $\nabla_\mu \Gg = \wt{\nabla}_\mu \Gf = 0$.

\section{Physical parameters}
\label{sec:PhysParam}
The $\beta$-parameters $\beta_0,...,\beta_4$ can be rescaled without affecting the physics. Hence, they are not observables and the reported constraints on them depend on the choice of scaling, being different in different papers. To circumvent this problem, a subset of the physical parameters was first introduced in Ref.~\cite{Luben:2020xll} and then generalized to the full theory in Ref.~\cite{PhysParamTh}. These parameters are independent of the rescaling and are thus observable quantities. The framework applies to space-times where the metrics are proportional asymptotically (in space or time), which we assume.

At radial infinity of static, spherically symmetric solutions (i.e., local solutions, applicable on for example solar system scales) the metrics are proportional, that is $\g = c^2 \f$ with $c=\mathrm{const}$. This is also true in the infinite future for the cosmological background solutions. We introduce dimensionless, rescaling invariant parameters \cite{PhysParamTh},
\begin{equation}
\label{eq:Bdef}
B_n \equiv \kappa_g \beta_n c^n/H_0^2,
\end{equation}
expressing the $\beta$-parameters in units of the curvature scale defined by the Hubble constant. This is especially convenient for cosmological application since we expect $\beta_n \sim H_0^2$, hence $B_n \sim 1$, if the theory is to exhibit novel features on cosmological scales. We can express the dimensionless physical parameters in terms of the $B$-parameters,
\begin{subequations}
	\label{eq:PhysToBeta}
	\begin{align}
	\tan^2 \theta &= \frac{B_1 + 3 B_2 + 3 B_3 + B_4}{B_0 + 3 B_1 + 3 B_2 + B_3},\\
	\mfphat^2 &= \left(B_1 + 2 B_2+ B_3\right) / \sin^2 \theta,\\
	\omegaleff &= \frac{B_0}{3} + B_1 + B_2 + \frac{B_3}{3},\\
	\alpha &= - \frac{B_2 + B_3}{B_1 + 2 B_2 + B_3},\\
	\beta &=\frac{B_3}{B_1 + 2 B_2 + B_3}.
	\end{align}
\end{subequations}
Inverting the relations,
\begin{subequations}
	\label{eq:BetaToPhys}
	\begin{align}
	\label{eq:B0phys}
	B_0 &= 3 \omegaleff - \sin^2 \theta \, \mfphat^2 (3 + 3\alpha + \beta),\\
	B_1 &= \sin^2 \theta \, \mfphat^2 (1+ 2 \alpha + \beta),\\
	B_2 &= - \sin^2 \theta \, \mfphat^2 (\alpha + \beta),\\
	B_3 &= \sin^2 \theta \, \mfphat^2 \beta,\\
	B_4 &= 3 \tan^2 \theta \, \omegaleff + \sin^2 \theta \, \mfphat^2 (-1 + \alpha - \beta).
	\end{align}
\end{subequations}
The physical parameters have specific physical interpretations: $\theta \in [ 0,\pi / 2]$ is the mixing angle between the mass eigenstates and the metrics (cf. the mixing between neutrino mass and flavor eigenstates), $\mfphat >0$ is the mass of the (massive) graviton measured in units of $H_0 = 100h \, \mathrm{km/s/Mpc} = 2.1h \times 10^{-33} \, \mathrm{eV}/c^2$, $\omegaleff>0$ is the effective cosmological constant in the infinite future, and $\alpha \in \mathbb{R}$ and $\beta \in \mathbb{R}$ determine (among other things) how and if the screening mechanism is active. The mixing angle can be expressed in terms of the ratio of the gravitational constants and the conformal factor,
\begin{equation}
\tan^2 \theta \equiv \kappa c^2.
\end{equation}
In the limit $\theta \to 0$, the physical metric coincides with the massless spin-2 field and we recover general relativity (GR). In the limit $\theta \to \pi / 2$, the physical metric coincides with the massive spin-2 field and we recover dRGT (de Rham--Gabadadze--Tolley) massive gravity \cite{deRham:2010ik,deRham:2010kj} with a fixed second metric, see for example \cite{Hassan:2012wr,Hassan:2014vja}.

\subsection{Analytical constraints}
\label{sec:AnConstr}
\begin{figure}[t]
	\centering
	\includegraphics[width=\linewidth]{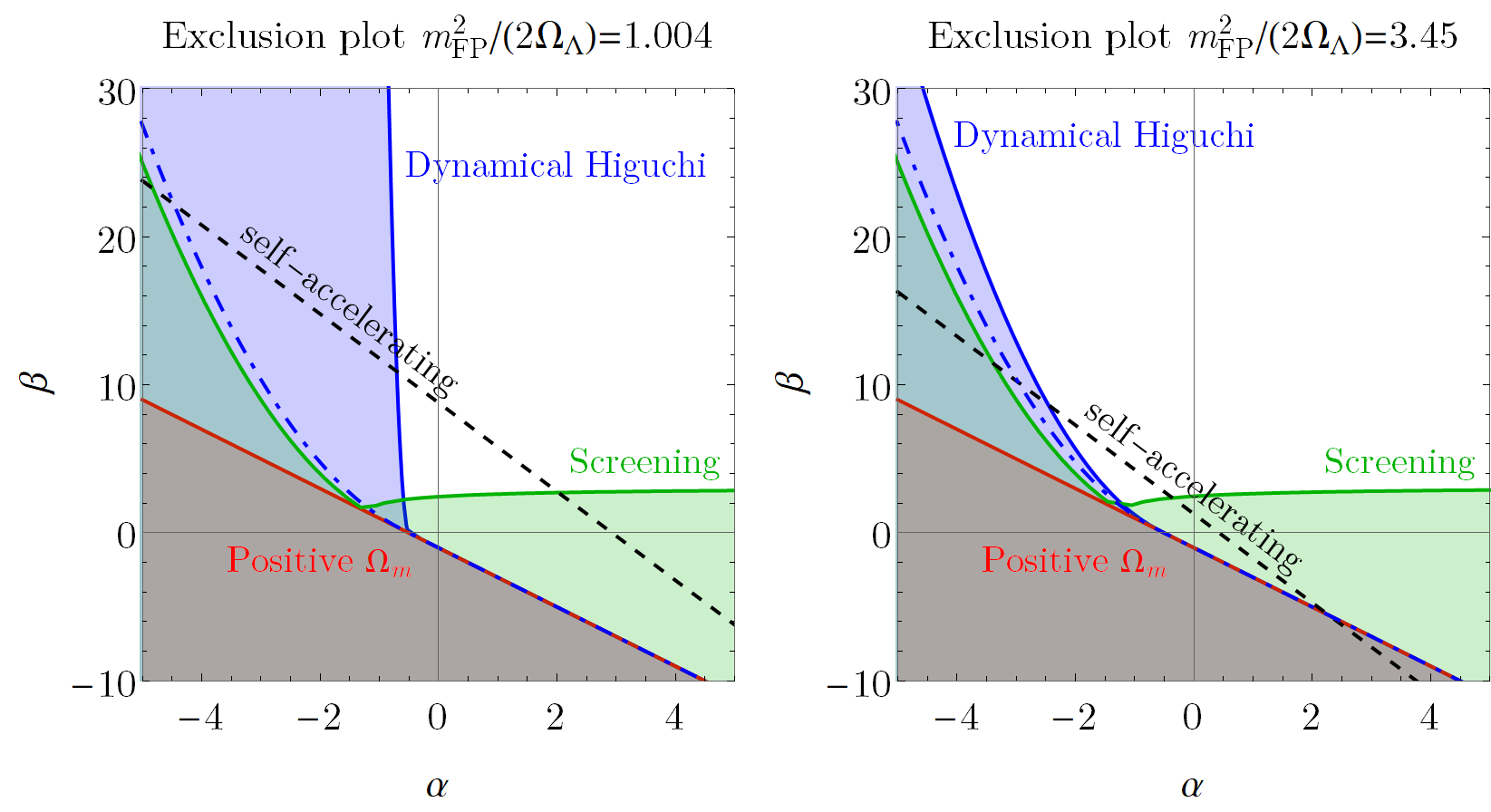}
	\caption{Exclusion plot in the $\alpha \beta$-plane from demanding a working screening mechanism (green), a continuous background cosmology without the Higuchi ghost (blue), and a positive matter density (red). Here, $\theta \simeq 20^\circ$. The dash-dotted curve indicates the boundary of the blue region in the limit $\mfphat^2 / 2 \omegaleff \to \infty$. The exclusion regions change only weakly with $\theta$. The blue region is affected by the value of $\mfphat^2 / 2 \omegaleff$ while the screening constraint (green) remains the same. The self-accelerating cosmological models lie along the dashed lines. The slope of these are always the same ($\beta = -3\alpha + \mathrm{const.}$) but the intersection with the vertical axis moves downwards if $\mfphat^2/(2\omegaleff)$ increases or if $\theta$ increases.}
	\label{fig:dynhig2}
\end{figure}

Here, we summarize the analytical constraints on the physical parameters presented in Ref.~\cite{PhysParamTh} and summarized in Fig.~\ref{fig:dynhig2}. To have a working screening mechanism restoring GR results on local length scales (e.g., in the solar system), we must impose constraints on $\alpha$ and $\beta$, the exact form depending on $\theta$ and $\mfphat$. To have a continuous, real-valued background cosmology devoid of the Higuchi ghost, we must put additional constraints on $\alpha$ and $\beta$, the exact form depending on $\theta$ and $\mfphat^2 / 2 \omegaleff$. It is possible to express these constraints in analytical form \cite{PhysParamTh}. In particular, all $B$-parameters (except $B_0$ and $B_4$) must be nonzero.

\subsection{Special models}
\label{sec:SpecModels}
In general, the physical parameters are independent. However, it is common to consider subsets of models where one or several of the $B$-parameters are set to zero, in which case the physical parameters are not independent \cite{Luben:2020xll}. To be consistent with the analytical constraints, $B_1$, $B_2$, and $B_3$ must all be non-vanishing, excluding all two-parameters models\footnote{The $B_0 B_4$ model (i.e., with only $B_0$ and $B_4$ non-vanishing) is still allowed. However, this is just two independent copies of general relativity.} (i.e., with all but two of the $B$-parameters set to zero). The possible submodels are: $B_1 B_2 B_3 B_4$ (self-accelerating), $B_0 B_1 B_2 B_3$, and $B_1 B_2 B_3$ (minimal model). Here, we do not discuss the $B_0 B_1 B_2 B_3$ model explicitly, for two reasons: (i) it does not give as good fit to data as the self-accelerating model which has the same number of free parameters (see Section~\ref{sec:Cosmo}) and (ii) it exhibits the cosmological constant $B_0$ which makes it less interesting from a theoretical perspective. To connect with earlier studies, although incompatible with the analytical constraints, we study some two-parameter models in Appendix~\ref{sec:TwoParam}.

\begin{figure}[t]
	\centering
	\includegraphics[width=0.9\linewidth]{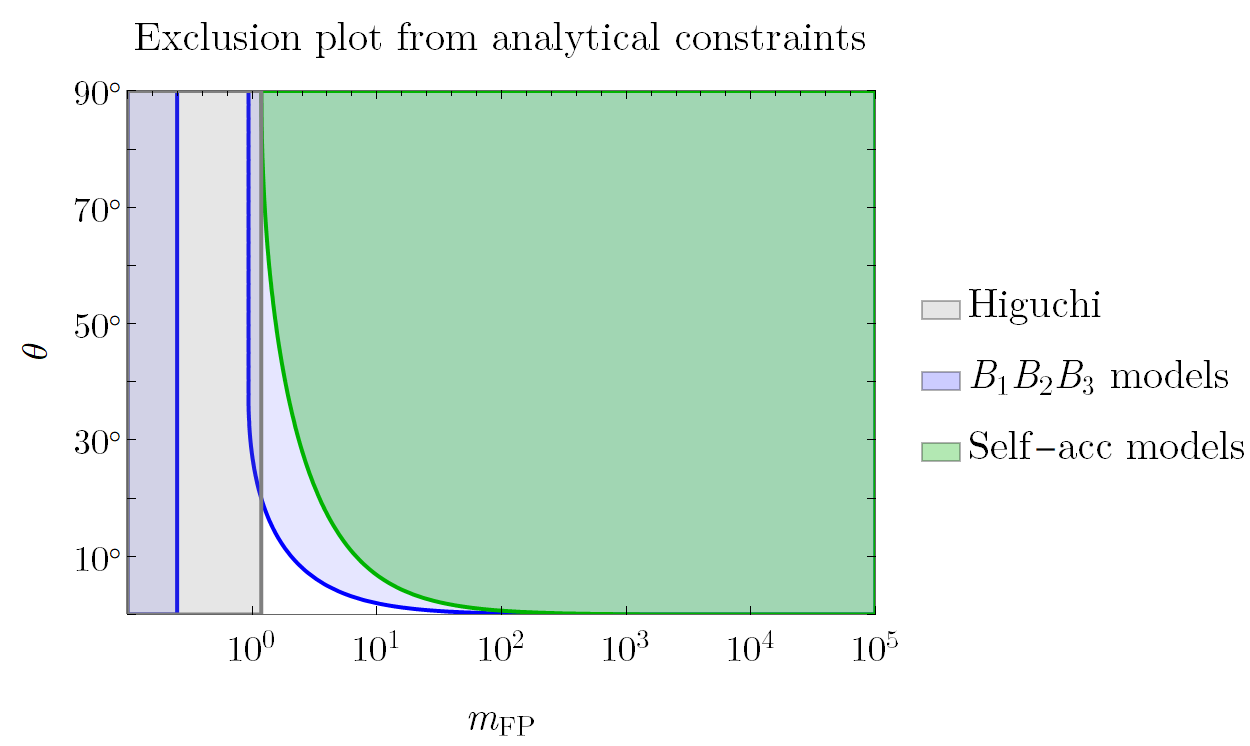}
	\caption{Exclusion plot due to the analytical constraints (i.e., requiring a working screening mechanism and a real-valued cosmology devoid of the Higuchi ghost). Here, we have set $\omegaleff = 0.7$. The green constraint applies to self-accelerating (i.e., $B_1 B_2 B_3 B_4$) models and the blue constraint to the $B_1 B_2 B_3$ (minimal) models. For these, the viable region is a limited area close to $\mfphat \sim 1$. Remember that $\mfphat$ is measured in units of $H_0$.}
	\label{fig:anconstrspecmodels}
\end{figure}

\paragraph{Self-accelerating models ($B_1 B_2 B_3 B_4$).}  Cases where $B_0 = 0$ are referred to as self-accelerating models and are of particular interest since there appears no cosmological constant term in the general equations of motion for the physical metric $\g$, see for example \cite{PhysParamTh}. Still, in the equations of motions for the background cosmology there appears an effective cosmological constant which is due to the interaction between the massless and massive spin-2 fields. Setting $B_0 =0$ in \eqref{eq:BetaToPhys}, we get a linear relation between $\alpha$ and $\beta$,
\begin{equation}
\label{eq:B0alphabeta}
\alpha + \frac{\beta}{3} = -1 + \frac{1}{\sin^2 \theta} \frac{\omegaleff}{\mfphat^2}, \quad B_0=0,
\end{equation}
that is, self-accelerating cosmologies lie along the line in the $\alpha \beta$-plane shown in Fig.~\ref{fig:dynhig2}. In order to have viable self-accelerating solutions satisfying the constraints introduced in Section~\ref{sec:BR}, $\theta$ is bounded from above. An approximate upper limit is obtained by requiring that the line \eqref{eq:B0alphabeta} lies above the point $(\alpha,\beta)=(-1/2,0)$, implying,
\begin{equation}
\label{eq:kappalimSA}
\sin^2 \theta \lesssim 2\omegaleff/ \mfphat^2, \quad B_0=0,
\end{equation}
which is an upper bound on $\theta$ for fixed values of $\mfphat^2 /(2\omegaleff)$. Together with the Higuchi bound $\mfphat^2 > 2 \omegaleff$, this leaves us with a viable region close to $\mfphat \sim 1$ unless $\theta$ is very small, see Fig.~\ref{fig:anconstrspecmodels}. Thus, due to the Higuchi bound, there is no dRGT massive gravity limit ($\theta \to \pi / 2$) for self-accelerating models. From eq. \eqref{eq:kappalimSA}, we also see that the infinite graviton mass limit $\mfphat \to \infty$ enforces the GR limit $\theta \to 0$ (for finite $\omegaleff$). Hence, for the self-accelerating models to have interesting cosmological background solutions (i.e., different from $\Lambda$CDM), $\mfphat$ cannot take too large values. From Fig.~\ref{fig:dynhig2}, we see that the limits $\alpha \to \infty$ and $\beta \to \infty$ take us out from the allowed region and are not consistent with self-accelerating cosmologies.

\paragraph{$B_1 B_2 B_3$ (minimal) models.} The $B_1 B_2 B_3$ models are self-accelerating models where the cosmological constant terms in the two metric sectors are set to zero, that is $B_0 = B_4 =0$, where $B_4$ is the cosmological constant of the $\f$ metric. This is the most minimal model compatible with a working screening mechanism and a real-valued, continuous cosmology without the Higuchi ghost. For these models, a small value of the effective dark energy density is technically natural in the sense of 't Hooft \cite{tHooft:1979rat}. Since $B_0=B_4=0$, we can use \eqref{eq:BetaToPhys} to express $\alpha$ and $\beta$ in terms of $\theta$ and $\mfphat^2/\omegaleff$,
\begin{subequations}
	\begin{alignat}{2}
	\label{eq:AlphaB123}
	\alpha &= \frac{- \mfphat^2 / \omegaleff + 3 \cot^2 \theta -3\tan^2 \theta}{4 \mfphat^2 / \omegaleff},& \quad B_0 &= B_4 =0\\
	\label{eq:BetaB123}
	\beta &= 3 \frac{-2 \mfphat^2 / \omegaleff + \csc^2 \theta + 3 \sec^2 \theta}{4 \mfphat^2 / \omegaleff},& \quad B_0 &= B_4 =0.
	\end{alignat}
\end{subequations}
A working Vainshtein mechanism requires $\beta \gtrsim 3$ (see Ref.~\cite{PhysParamTh}), hence,
\begin{equation}
\label{eq:AnConstrB123}
\tan^2 \theta \lesssim \frac{1}{3} \left(-2 + 3 \mfphat^2 / \omegaleff - \sqrt{1 + 6 \mfphat^2 /\omegaleff (-2+3 \mfphat^2 / 2 \omegaleff)}\right),
\end{equation}
see Fig.~\ref{fig:anconstrspecmodels}. The largest value of $\theta$ is obtained by setting $\mfphat^2 = 2 \omegaleff$ (from the Higuchi bound, see \cite{PhysParamTh}). The result is,
\begin{equation}
\label{eq:ThetaUpperBound}
\theta \lesssim 20^{\circ}.
\end{equation}
For the mixing angle not be close to the GR limit, the ratio $\mfphat^2 /2 \omegaleff$ must be close to the Higuchi bound. This is indeed the case, see Tab.~\ref{tab:cosmoconstr}.

\newpage \section{Local tests (stars, galaxies, and galaxy clusters)}
Static, spherically symmetric (SSS) solutions can be used to approximate the gravitational potentials of for example the solar system, galaxies, and galaxy clusters. We refer to them as local solutions as opposed to cosmological solutions. In Refs. \cite{Enander:2015kda,Luben:2018ekw}, constraints were placed on the theory parameters assuming $\alpha \sim \beta \sim 1$. In Ref.~\cite{Platscher:2018voh}, the authors constructed a phenomenological model, parameterized by the mixing angle $\theta$ and the graviton mass $\mfphat$. Here, we consider a general bimetric model including $\alpha$ and $\beta$. In principle, the effective cosmological constant $\omegaleff$ dictates the asymptotic structure at radial infinity. However, local solutions are not sensitive to the value of $\omegaleff$ which becomes influential only when approaching the Hubble length scale, so $\omegaleff$ does not affect the results of this section.

There are two types of approximate analytical solutions to the SSS equations of motion, applicable under different circumstances: the linearized solutions and the nonlinear Vainshtein screening solutions. A conservative estimate is that the linearized solutions are compatible with solar system data only if $\mfp \gtrsim 10^{33}$ or $\theta \lesssim 10^{-5}$ \cite{Luben:2018ekw}. In both cases, the background cosmology reduces to $\Lambda$CDM for all practical purposes and hence we loose many of the interesting features of bimetric gravity \cite{PhysParamTh}.\footnote{To be precise, the background cosmology of the large graviton mass limit reduces to $\Lambda$CDM, provided that we do not impose any independent observation of the dimensionless matter density $\omegamnot$. If such a value is imposed, we also need to choose $\theta \to 0$ to be observationally viable, again pushing the model to its GR limit.} Therefore, to ensure that our models are compatible with local tests of gravity while at the same time having novel cosmological solutions, we demand the existence of a working Vainshtein mechanism. It should be stressed however, that a careful analysis may reveal regions in the parameter space in which the linearized solutions pass the local tests even if $\mfphat < 10^{33}$ or $\theta > 10^{-5}$.

The Vainshtein screening mechanism restores general relativity inside some radius, provided that we satisfy the analytical constraints of Section~\ref{sec:AnConstr}; $\alpha$ and $\beta$ determine how the screening mechanism is realized (together with $\theta$, $\mfphat$, and the mass of the source) \cite{Vainshtein:1972sx,Babichev:2013pfa,Enander:2015kda,PhysParamTh}. The radius within which GR can be restored is proportional to the Vainshtein radius $r_V$,
\begin{equation}
	r_V \equiv \left(2M / H_0^2 \mfphat^2 \right)^{1/3},
\end{equation}
where $M$ is the mass of the source. In the regions far inside and far outside the Vainshtein radius, the gravitational potential takes a simple form,
\begin{subequations}
	\begin{alignat}{2}
	\Phi &= -\frac{2M}{r},& \quad r \ll r_V,\\
	\Phi &= - \frac{2M}{r} \left(1+\frac{1}{3} \sin^2 \theta \right),& \quad r \gg r_V,
	\end{alignat}
\end{subequations}
and in the intermediate regions, it is calculated according to Ref.~\cite{PhysParamTh}. If $\alpha \sim \beta \sim 1$, the Vainshtein radius sets the length scale within which we start to approach GR. As an example, to have a viable background cosmology, typically $\mfphat \sim 1$,  and hence,
\begin{equation}
r_V / r_* \sim \left(\rho_* / \rho_c \right)^{1/3},
\end{equation}
where $r_*$ is the radius of the source and $\rho_*$ is the mean density. Hence, the Vainshtein mechanism is relevant for all objects with a density greater than the critical density $\rho_c \equiv 3H_0^2 / \kappa_g$ and of course also all astronomical objects like the solar system, galaxies, and galaxy clusters, invalidating the applicability of linear structure formation severely for bimetric gravity \cite{Mortsell:2015exa}.\footnote{The critical density is the total density (excluding curvature $\Omega_k$) that the Universe must have today in order to be spatially flat/Euclidean.}  For the Sun, the Vainshtein radius is $r_V \sim 10^7 \,  \mathrm{AU}$ where the gravitational force is negligible anyway. For the Milky way, $r_V^\mathrm{Milky \, Way} \sim 100 \, r_*$ where we used $M \sim 10^{12}M_\odot$ and $r_* \sim 50 \, 000 \, \mathrm{ly}$.

\begin{figure}[t]
	\centering
	\includegraphics[width=0.9\linewidth]{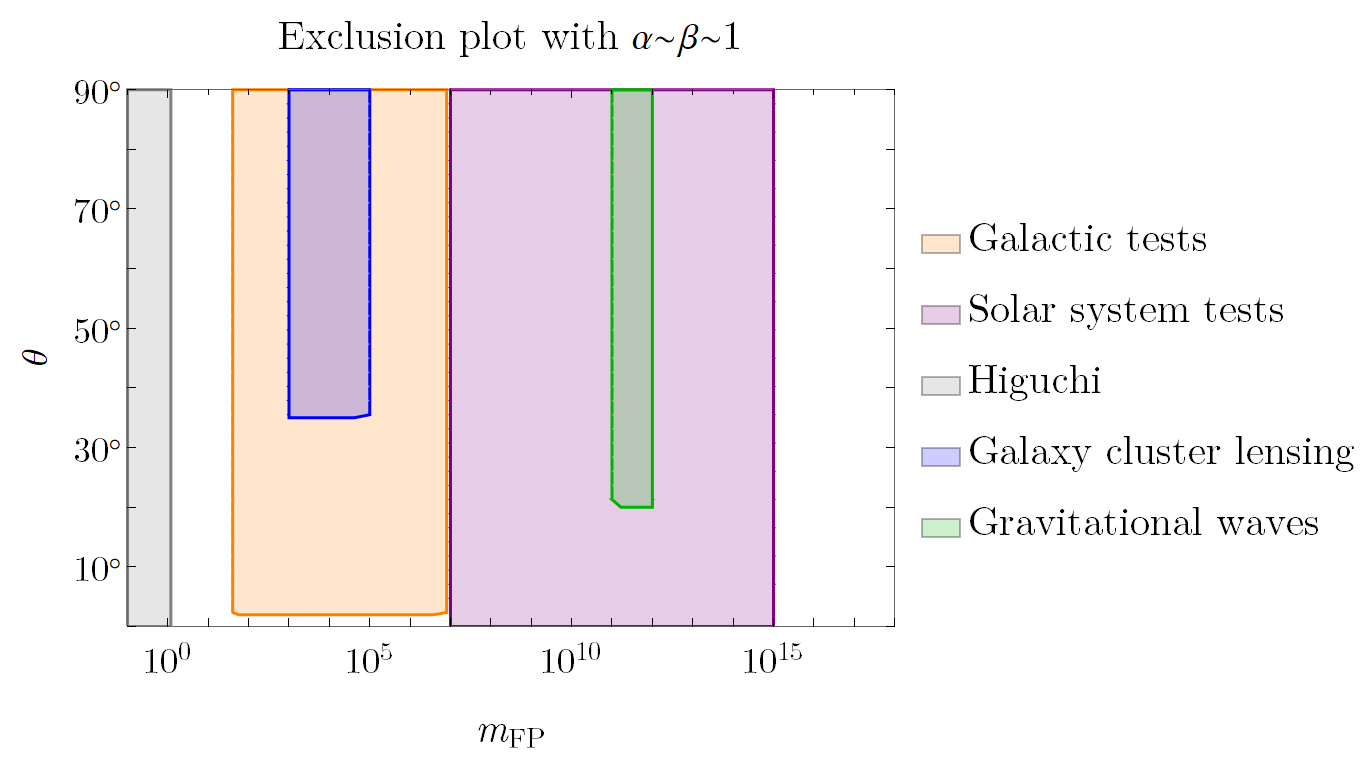}
	\caption{Exclusion plot, assuming $\alpha \sim \beta \sim 1$. Order of magnitude observational constraints from galactic tests (orange) \cite{Enander:2015kda}, solar system tests (purple) \cite{Luben:2018ekw}, galaxy cluster lensing (blue) \cite{Platscher:2018voh}, and gravitational waves (green) \cite{Max:2017flc}. The Higuchi bound is shown in gray (here, we have set $\omegaleff = 0.7$) \cite{Higuchi:1986py}. Remember that $\mfphat$ is measured in units of $H_0$.}
	\label{fig:localtestscompiled}
\end{figure}

If we push $\alpha$ or $\beta$ away from unity, we increase the radius within which we start to approach GR \cite{PhysParamTh}. Hence, $\alpha \to \infty$ and $\beta \to \infty$ are GR limits for the local Vainshtein screening solutions. This is true for general bimetric models where all the physical parameters are independent although, as discussed, inconsistent for self-accelerating models. The cases $\theta \to 0$ and $\mfphat \to 0$ are also GR limits. However, the latter limit is problematic since it excites the Higuchi ghost. Since the local solutions approach GR in the large parameter limits of $\alpha$ and $\beta$, the constraints placed on $\theta$ and $\mfphat$, assuming $\alpha \sim \beta \sim 1$ (e.g. in Refs. \cite{Enander:2015kda,Platscher:2018voh,Luben:2018ekw}), are alleviated in these limits. In Fig.~\ref{fig:localtestscompiled} we show the observational constraints from local tests of gravity, assuming $\alpha \sim \beta \sim 1$, together with the Higuchi bound. Comparing Figs.~\ref{fig:anconstrspecmodels} and \ref{fig:localtestscompiled}, we see that self-accelerating models (including the $B_1 B_2 B_3$ models) which satisfy the analytical constraints automatically pass the local tests of gravity. The observational constraints that we use in this section are order of magnitude estimates. Detailed results can be found in the corresponding references.

\newpage \subsection{Solar system tests}
\begin{figure}[t]
	\centering
	\begin{subfigure}[b]{0.49\textwidth}
		\centering
		\includegraphics[width=\textwidth]{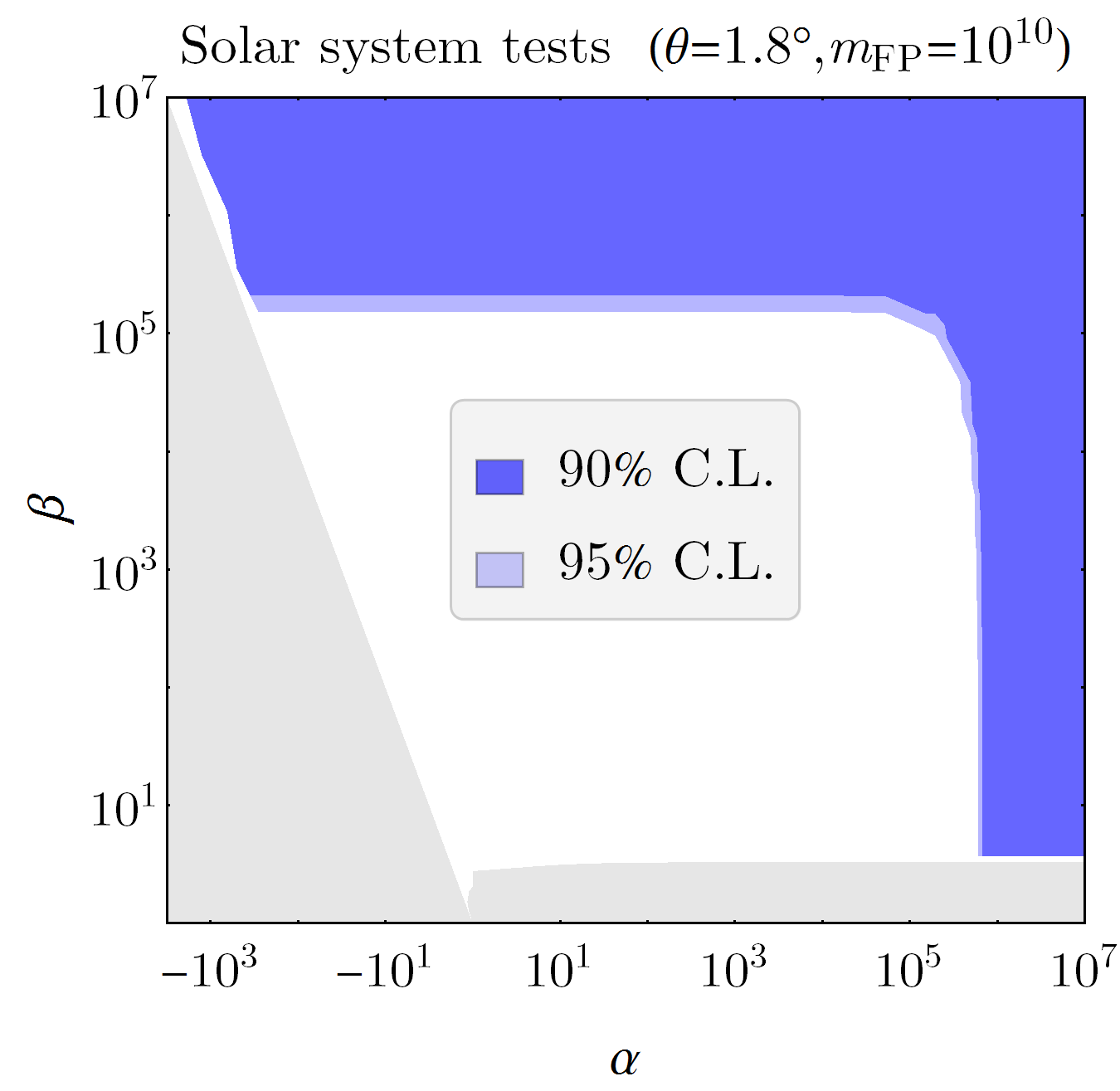}
	\end{subfigure}
	\hfill
	\begin{subfigure}[b]{0.49\textwidth}
		\centering
		\includegraphics[width=\textwidth]{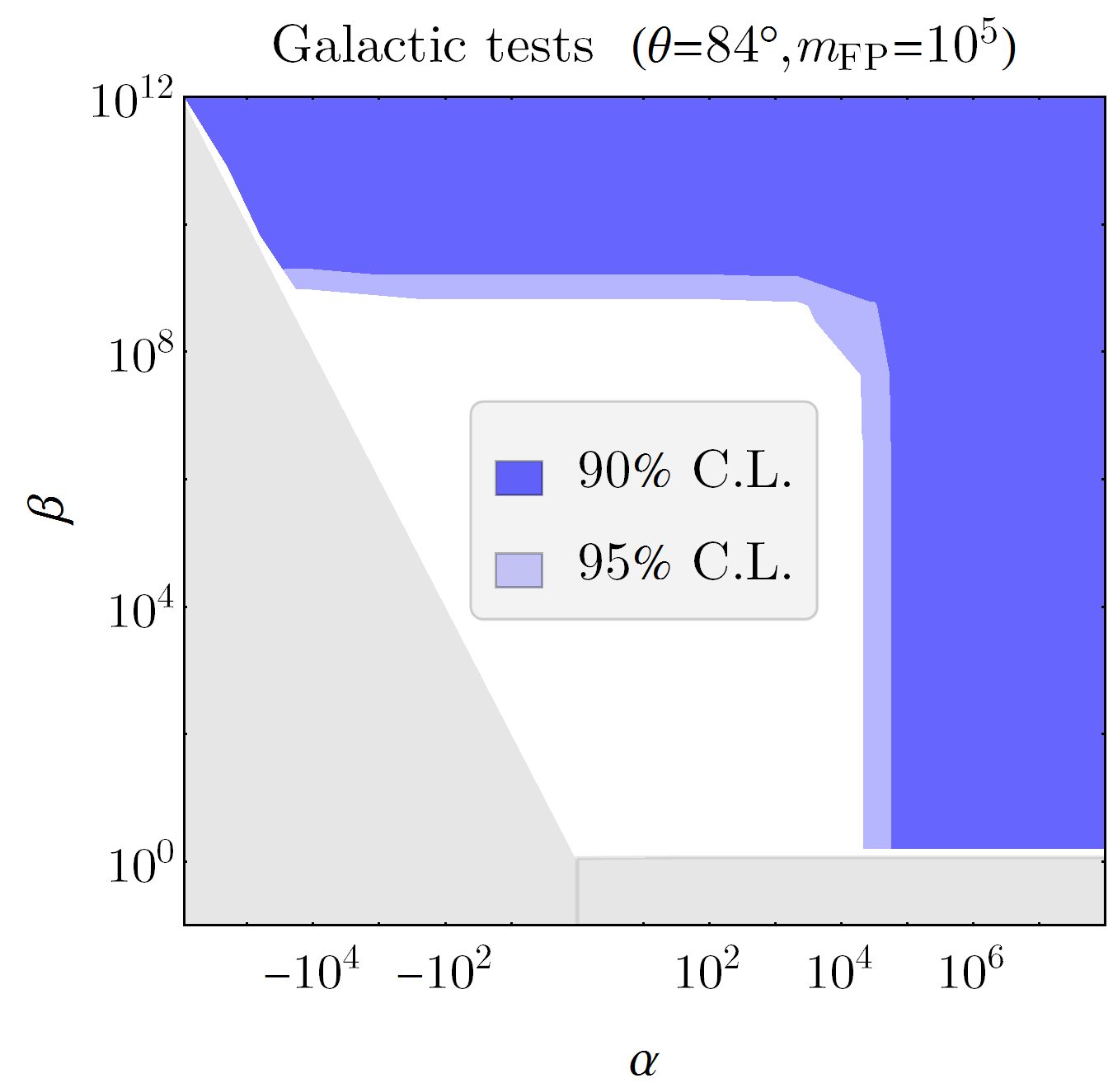}
	\end{subfigure}
	\caption{Confidence contours in the $\alpha\beta$-plane. The effective cosmological constant does not appear in the local solutions. For large $\alpha$ or $\beta$, GR is restored. The gray region is excluded by the requirement of a working Vainshtein mechanism, cf. Fig.~\ref{fig:dynhig2}. \emph{Left panel}: Solar system tests with $\theta=1.8^{\circ}$ and $\mfphat=10^{10}$. \emph{Right panel}: Gravitational lensing by a $10^{11} \, M_\odot$ galaxy. Here, $(\theta = 84^{\circ},\mfphat=10^5)$.}
	\label{fig:localtests}
\end{figure}

A conservative estimate is that bimetric gravity is compatible with solar system tests (at $1 \, \mathrm{AU}$) except if $\theta \gtrsim 10^{-5}$ in the range $10^{7} \lesssim \mfphat \lesssim 10^{15}$ \cite{Luben:2018ekw}. The possibility of alleviating the constraints on $\theta$ and $\mfphat$ by large values of $\alpha$ or $\beta$ is illustrated in Fig.~\ref{fig:localtests}. There, we plot confidence contours in the $\alpha \beta$-plane with $\theta = 1.8^{\circ}$ and $\mfphat = 10^{10}$ which would be observationally excluded if $\alpha \sim \beta \sim 1$. As a rough estimate, solar system tests constrain the gravitational force to be proportional to $1/r^2$ to an accuracy of $10^{-9}$ \cite{Will:2014kxa}. Hence, we calculate the chi-squared values as,
\begin{equation}
\chi^2_\mathrm{solar \; system} = \left. \left(\frac{(\Phi_\mathrm{GR}-\Phi)/\Phi_\mathrm{GR}}{10^{-9}}\right)^2 \right|_{r=1 \, \mathrm{AU}}.
\end{equation}
Here, $\Phi_\mathrm{GR} = -2M/r$. The confidence contours are surface levels of $\Delta \chi^2 \equiv \chi^2 - \chi^2_\mathrm{min}$ where the $90 \, \%$ level is at $\Delta \chi^2 = 4.61$ and the $95 \, \%$ level is at $\Delta \chi^2 = 5.99$. For the best-fit cosmological parameters (see Section~\ref{sec:Cosmo}), the relative difference between $\Phi$ and $\Phi_\mathrm{GR}$ is $\simeq 10^{-30}$, so it is well within the observational constraints.

\subsection{Galactic tests}
As a rough model of gravitational lensing on galactic scales, we fix the galaxy mass and calculate the lensing radius using the equation (in geometrized units),
\begin{equation}
r_\mathrm{lens} = 4 M D_\mathrm{source}/R,
\end{equation}
where $D_\mathrm{source}$ is the distance of the source (which is set to $1 \, \mathrm{Gpc}$). For a galaxy of mass $M = 10^{11} M_\odot$, the lensing radius is $r_\mathrm{lens} \simeq 8 \, \mathrm{kpc}$.  For simplicity, we assume that the lensing takes place at the edge or outside of the galaxy.

\noindent Gravitational lensing together with dynamical measurements of the mass probe the difference in the potential felt by the massive and massless particles, that is $\Phi$ and $\varphi \equiv (\Phi + \Psi)/2$, respectively. Here, $\Phi$ is the gravitational potential and $\Psi$ is the scalar curvature. The gravitational slip is defined as,
\begin{equation}
\gamma = \Phi / \varphi.
\end{equation}
We conservatively assume that measurements constrain $\gamma$ to be within $10 \, \%$ of the GR value $\gamma_\mathrm{GR}=1$ \cite{Schwab:2009nz}, calculating the chi-squared values as,
\begin{equation}
\chi^2_\mathrm{gal.lens.} = \left(\frac{\gamma(r_\mathrm{lens})-1}{0.1}\right)^2.
\end{equation}
Assuming $\alpha \sim \beta \sim 1$, this constrains $\theta \lesssim 2^{\circ}$ in the graviton mass range $40 \lesssim \mfphat \lesssim 8 \times 10^6$ \cite{Enander:2015kda}. Increasing $\alpha$ and $\beta$, these constraints are alleviated, see Fig.~\ref{fig:localtests}. For the best-fit cosmological parameters (see Section~\ref{sec:Cosmo}), the deviation of the gravitational slip from unity is at the level $10^{-4}$ which is well within the theoretical constraints.

\subsection{Cluster lensing}
In Ref.~\cite{Platscher:2018voh}, the authors considered a phenomenological model for the local solutions, parameterized by the mixing angle and the graviton mass. Comparing with clustering data constrains $\theta \lesssim 35^{\circ}$ in the graviton mass range $10^3 \lesssim \mfp \lesssim 10^5$. As in the case of galactic tests, the constraint can be alleviated by pushing $\alpha$ or $\beta$ to large values.

\section{Cosmological tests}
\label{sec:Cosmo}
\subsection{The modified Friedmann equation}
\begin{figure}[t]
	\centering
	\includegraphics[width=0.8\linewidth]{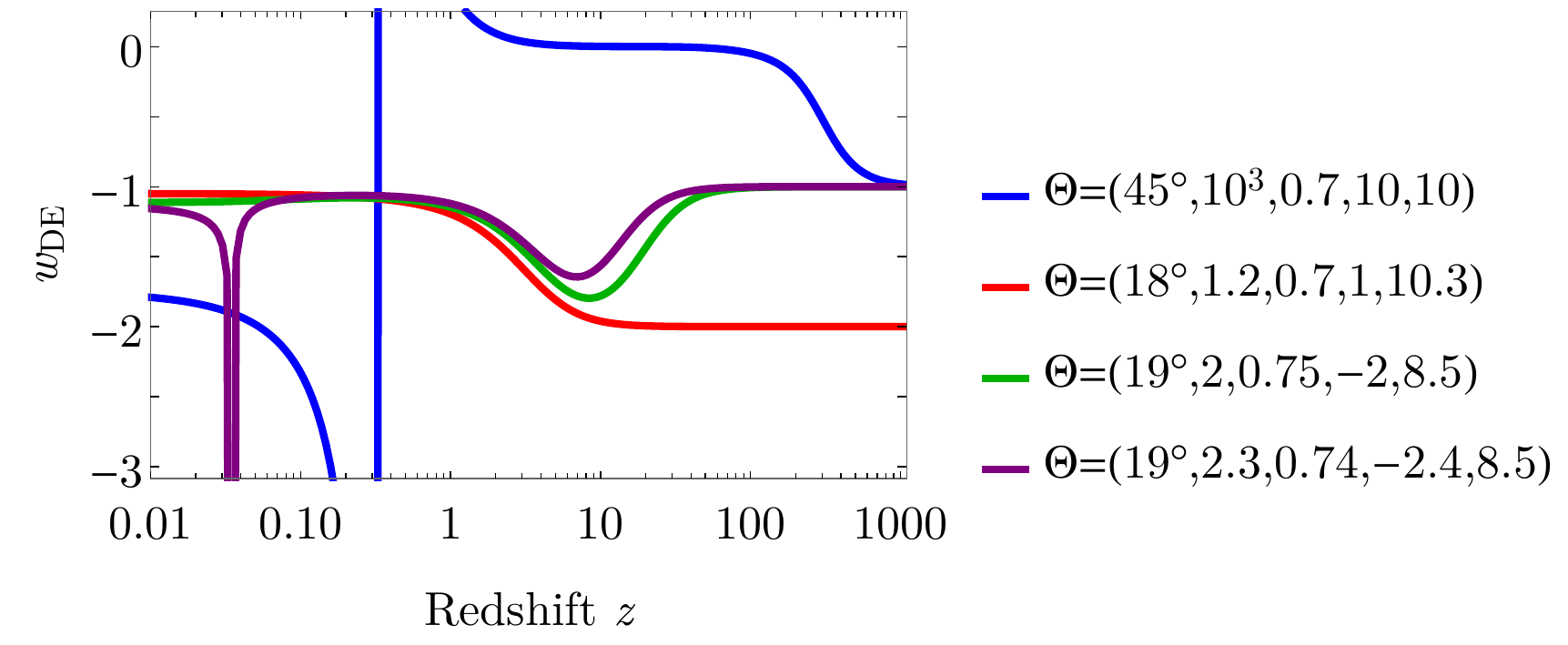}
	\caption{Examples of the equation of state (EoS) for the bimetric dark energy fluid, for different physical parameters $\Theta = (\theta,\mfphat,\omegaleff,\alpha,\beta)$. The red curve is a self-accelerating model. All the models approach $w_\mathrm{DE}=-1$ in the infinite future. In the early universe they also approach $w_\mathrm{DE}=-1$, except self-accelerating models which approach $w_\mathrm{DE}=-(2+w_m)$. In the intermediate region where the massive spin-2 field is dynamical, the equation of state depends on the physical parameters. For the blue curve, $\omegade$ is negative in the early universe, hence it diverges when crossing zero at $z \simeq 0.3$. Note that there is no physical singularity at this point. For the blue curve, the massive spin-2 field contributes as a dark matter component (i.e., $w_\mathrm{DE}=0$) in the redshift range $3 \lesssim z \lesssim 100$. The red curve is a self-accelerating model and has a smooth transition from $w_\mathrm{DE}=-2$ to $w_\mathrm{DE}=-1$ in the range $1 \lesssim z \lesssim 10$. For the green curve, the bimetric fluid contributes with a cosmological constant term except in the range $1 \lesssim z \lesssim 100$ where it has a phantom dark energy phase. The purple model is similar to the green one, with the difference that it also has a very brief phantom phase at $z \simeq 0.03$. Note that equation of state does not diverge here but has a rather sharp dip reaching a minimum of $w_{\mathrm{DE,min}}\simeq-12$.}
	\label{fig:wde}
\end{figure}

Since we have two metrics, the geometry of a homogeneous and isotropic universe is described by two scale factors: $a(t)$ for the physical metric $\g$ and $\wt{a}(t)$ for the second metric $\f$. In the asymptotic future, the metrics tend to a proportional de Sitter solution $\f = c^2 \g$ ($c=\mathrm{const}$). Instead of the scale factor $\wt{a}$, it is convenient to use the ratio between the scale factors (including $c$ and omitting the argument $t$),
\begin{equation}
	y \equiv \wt{a}/(ca),
\end{equation}
which can be expressed in terms of the matter density $\Omega_m$ by solving the quartic polynomial,
\begin{align}
\label{eq:yPoly}
& - \frac{1}{3} \cos^2 \theta \, \mfphat^2 (1+2\alpha+\beta) + \left[\Omega_m + \omegaleff + \mfphat^2 \left(\cos^2 \theta \, (\alpha + \beta) - \sin^2 \theta \, \left(1+\alpha + \frac{\beta}{3}\right)\right)\right] y \nonumber \\ 
& + \mfphat^2 \left[-\cos^2 \theta \, \beta + \sin^2 \theta \, (1+2\alpha+\beta)\right] y^2 \nonumber\\
&-\left[\omegaleff  + \frac{1}{3} \mfphat^2 \left(\cos^2 \theta \, (-1+\alpha-\beta) + 3\sin^2 \theta \, (\alpha+\beta)\right)\right] y^3 + \frac{1}{3} \sin^2 \theta \, \mfphat^2 \beta y^4 = 0.
\end{align}
The modified Friedmann equation for the scale factor $a(t)$ reads (omitting the argument $t$),
\begin{equation}
\label{eq:BRFriedm}
	E^2 = \Omega_m + \Omega_k + \omegade, \quad E \equiv H / H_0.
\end{equation}
Here, $H = \dot{a}/a$ is the Hubble parameter of the physical metric and $H_0$ is the Hubble parameter evaluated today (i.e., the Hubble constant). $\omegade$ and $\Omega_m$ are the dimensionless energy densities measured in units of the critical energy density today, $\rho_c \equiv 3 H_0^2 / \kappa_g$ and $\Omega_k$ is the contribution from spatial curvature,
\begin{equation}
		\Omega_m = \frac{\rho_m}{\rho_c} = \Omega_{m,0} (1+z)^{3(1+w_m)}, \quad \Omega_k = - \frac{k}{H_0^2 a^2} = \Omega_{k,0} (1+z)^2,
\end{equation}
with $\rho_m$ being the physical matter energy density. For simplicity, we consider only matter content with the same equation of state $w_m$ in eq. \eqref{eq:BRFriedm}. However, we can generalize straightforwardly by adding new terms for additional matter fields. Here, we are mostly interested in redshifts $0 \leq z \lesssim 1100$ and therefore set $w_m=0$ (pressureless dust) in this range. In the limit $z \to \infty$, we set $w_m = 1/3$ (radiation). $\Omega_\mathrm{DE}$ is a dynamical ``dark energy" contribution due to the gravitational interaction of the massive spin-2 field and can be expressed in terms of $y$ as,
\begin{equation}
\label{eq:OmegaDefs}
	\omegade = \omegaleff - \sin^2 \theta \, \mfphat^2 (1-y) \left[1 + \alpha (1-y) + \frac{\beta}{3} (1-y)^2\right].
\end{equation}
\begin{figure}[t]
	\centering
	\includegraphics[width=0.9\linewidth]{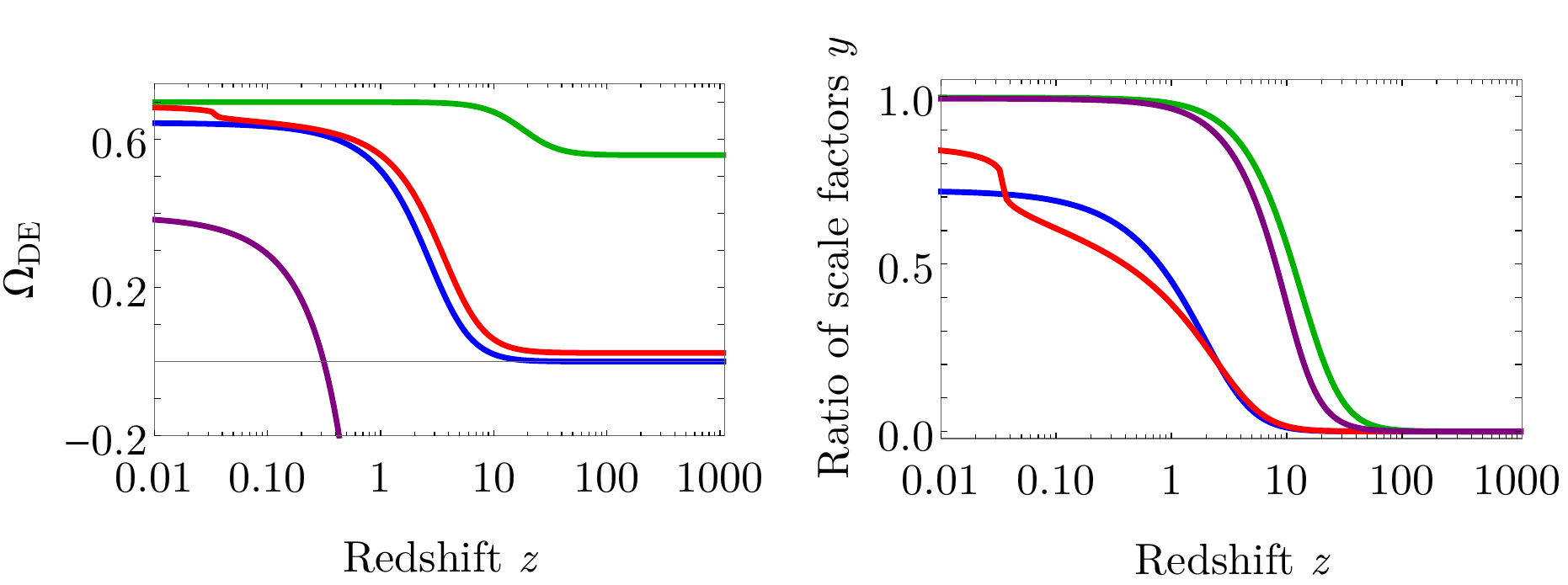}
	\caption{Examples of $\omegade(z)$ and $y(z)$ for different physical parameters, $\Theta = (\theta,\mfphat,\omegaleff,\alpha,\beta)$. Blue curve: $\Theta \simeq (18^{\circ},1.2,0.7,1,10.3)$. Red curve: $\Theta \simeq (18^{\circ},2.3,0.74,-2.4,8.5)$. Green curve: $\Theta \simeq (0.6^{\circ},10,0.70,10,10)$. Purple curve: $\Theta \simeq (45^{\circ},10,0.70,10,10)$. $\omegade$ and $y$ decrease monotonically with redshift. The blue curve is a self-accelerating model, hence the dark energy density is zero in the early universe (i.e., as $z \to \infty$). The red and green curves are models with positive dark energy density as $z \to \infty$ while the purple model has negative dark energy density in the $z \to \infty$ limit. The late-time cosmological constant is set by $\omegaleff$. In the limit $z \to \infty$, the dark energy density for the purple model approaches $B_0 / 3 \simeq -716$.}
	\label{fig:omegade_y}
\end{figure}
Since the equation for $y$ \eqref{eq:yPoly} is a quartic polynomial, it has a closed-form solution with up to four real solutions. However, only one of them is well-behaved. This is the finite/expanding branch solution which is defined as the lowest lying, strictly positive root of $y$ \cite{Luben:2020xll}. For these solutions, $y$ and $\omegade$ increase monotonically with time (i.e., decrease with redshift), with $y$ starting at $y|_{z=\infty}=0$ at the Big Bang and ending up at $y|_{z=-1}=1$ in the infinite future. The dark energy $\omegade$ starts at some value at the Big Bang (negative, positive, or zero depending on the value of the physical parameters) and ends up at the (positive) value $\omegaleff$ in the infinite future, see Fig.~\ref{fig:omegade_y} for some examples. For a general bimetric model, the equation of state for the dark energy, $w_\mathrm{DE}$, starts at $w_\mathrm{DE}|_{z=\infty}=-1$ in the early universe and ends up also at $w_\mathrm{DE}|_{z=-1}=-1$ in the future infinity. (For a self-accelerating model, $w_\mathrm{DE}|_{z=\infty}=-(2+w_m)$.) Hence, the bimetric fluid acts as a cosmological constant both in the early and late universe. However, since $\omegade$ is increasing with time, in the late universe it has a larger value than in the early universe, see for example Fig.~\ref{fig:omegade_y}. In the intermediate region between the early-time and late-time cosmological constant phases, there is a dynamical phase where the massive spin-2 field is dynamical, which can give rise to many different scenarios, depending on the values of the physical parameters, see Fig.~\ref{fig:wde} for some examples. Typically, in the intermediate phase, the effective dark energy fluid behaves as phantom dark energy. Since $w_\mathrm{DE} \to -1$ fast enough in the late universe, there is no Big Rip, which can be shown analytically, see Appendix E of \cite{PhysParamTh}.

\subsection{Fitting to data}
To explore the observational viability and to find the best fit parameters, we construct a discrete grid in the parameter space $\Theta$. For each point on the grid, we asses whether the analytical constraints of Section~\ref{sec:AnConstr} are satisfied. If not, the likelihood is set to zero. Thereby, we guarantee a working screening mechanism and that the cosmology is continuous, real-valued, and devoid of the Higuchi ghost. If the point is not excluded by these constraints, the likelihood is calculated by fitting the cosmological model to data from CMB, BAO, SNIa, and a background independent measurement of $\omegamnot$. Since there is no established framework for treating structure formation in bimetric theory, we will be conservative and combine CMB and BAO data in a way which effectively cancels the dependence on the cosmology before $z_* \simeq 1090$. We fit to the ratio of the cosmological distance to each of the BAO points and the comoving angular diameter distance to the last scattering surface at $z_* \simeq 1090$. The BAO data sets that we use are 6dFGS \cite{Beutler_2011}, SDSS MGS \cite{Ross:2014qpa}, BOSS DR12 \cite{Alam:2016hwk}, BOSS DR14 \cite{Bautista:2017wwp}, and eBOSS QSO \cite{Zhao:2018gvb}, that is, in total ten points in the redshift range $z \in [0.106,1.944]$. From CMB, we use the shift parameter $\ell_A$ from Planck 2018 \cite{Aghanim:2018eyx} as calculated in \cite{Chen:2018dbv}. For the type Ia supernovae, we use the binned Pantheon data set; 40 bins in the range $z \in [0.014,1.61]$ \cite{Scolnic:2017caz}. In Appendix~\ref{sec:DataSets}, we describe the details of the data sets and how the likelihood is computed. We also impose a background independent measurement of the matter density today, which is based on observations of the X-ray gas fraction in galaxy clusters (XCL) \cite{Allen:2004cd} and the extragalactic dispersion of fast radio bursts \cite{Macquart2020},
\begin{equation}
\label{eq:MatterPrior}
	\omegamnot = 0.29 \pm 0.09.
\end{equation}
As shown in Ref. \cite{PhysParamTh}, in the large parameter limits $\mfphat,\alpha,\beta \to \infty$, the current matter density is given by,
\begin{equation}
\label{eq:OmegaM0lims}
	\omegamnot = (1-\omegaleff)/\cos^2 \theta, \quad \mfphat,\alpha,\beta \to \infty.
\end{equation}
Hence, assuming $\omegaleff \simeq 0.7$, eq. \eqref{eq:OmegaM0lims} together with \eqref{eq:MatterPrior} set an upper limit on $\theta$ in these large parameter limits. Observations support $\Omega_k \simeq 0$, not only for a $\Lambda$CDM model but also for bimetric models \cite{Lindner:2020eez}. Therefore, we assume a spatially flat universe with $\Omega_k = 0$, leaving us with the bimetric physical parameters and $\Omega_{m,0}$ as cosmological parameters. (Since we only need the expansion history up to $z_* \simeq 1090$, we consider only pressureless dust with $w_m=0$.) Evaluating the modified Friedmann equation \eqref{eq:BRFriedm} and the $y$ polynomial \eqref{eq:yPoly} today, we get two equations, containing $\omegamnot$, $y_0$ and the physical parameters. We solve the equations for $y_0$ and $\Omega_{m,0}$, leaving only the bimetric physical parameters independent. Hence, when fitting to cosmological data we are directly exploring the likelihood of the physical parameter space.

\begin{figure}[t]
	\centering
	\includegraphics[width=0.65\linewidth]{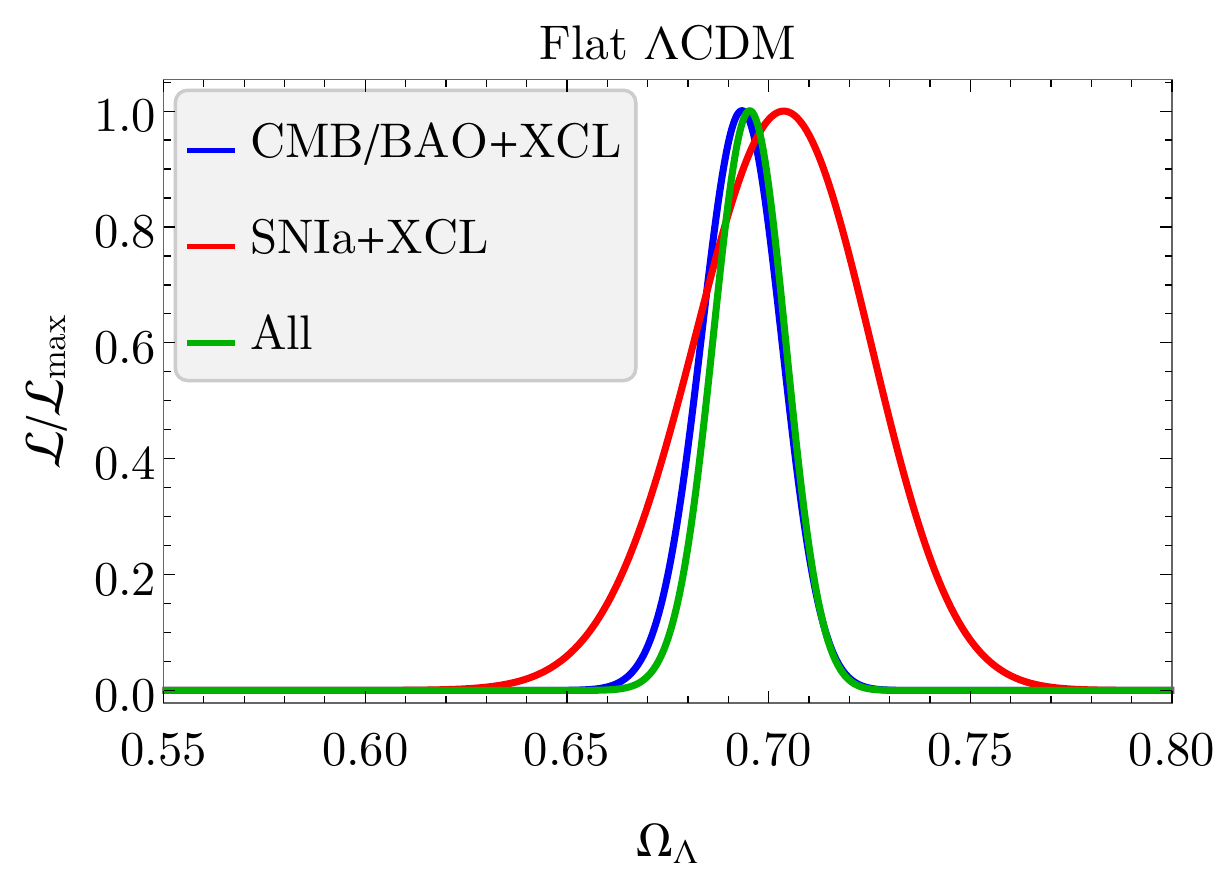}
	\caption{Normalized likelihood distribution for spatially flat $\Lambda$CDM models as a function of the cosmological constant $\omegal$. The best-fit point for CMB/BAO+XCL is $\omegal = 0.694$, and for SNIa+XCL, the best-fit is $\omegal = 0.704$. For all data combined, $\omegal = 0.695$ for which $\chi^2_\mathrm{min}=52.5$.}
	\label{fig:lcdmlikelihood}
\end{figure}

\subsection{Results}
\paragraph{$\Lambda$CDM.} In the spatially flat $\Lambda$CDM models with current matter density $\omegamnot = 1- \omegaleff$, the cosmological constant $\omegal$ is the only independent parameter. The result of fitting to CMB/BAO, SNIa, and $\omegamnot$ (from XCL) data is presented in Fig.~\ref{fig:lcdmlikelihood}. The value is $\omegal = 0.695 \pm 0.014$ (corresponding to $\omegamnot = 0.305 \pm 0.014$) for which $\chi^2_\mathrm{min} = 52.5$ (51 data points; 40 from SNIa, 10 from CMB/BAO, and one from XCL).

\begin{figure}[t]
	\centering
	\begin{subfigure}[b]{0.49\textwidth}
		\centering
		\includegraphics[width=\textwidth]{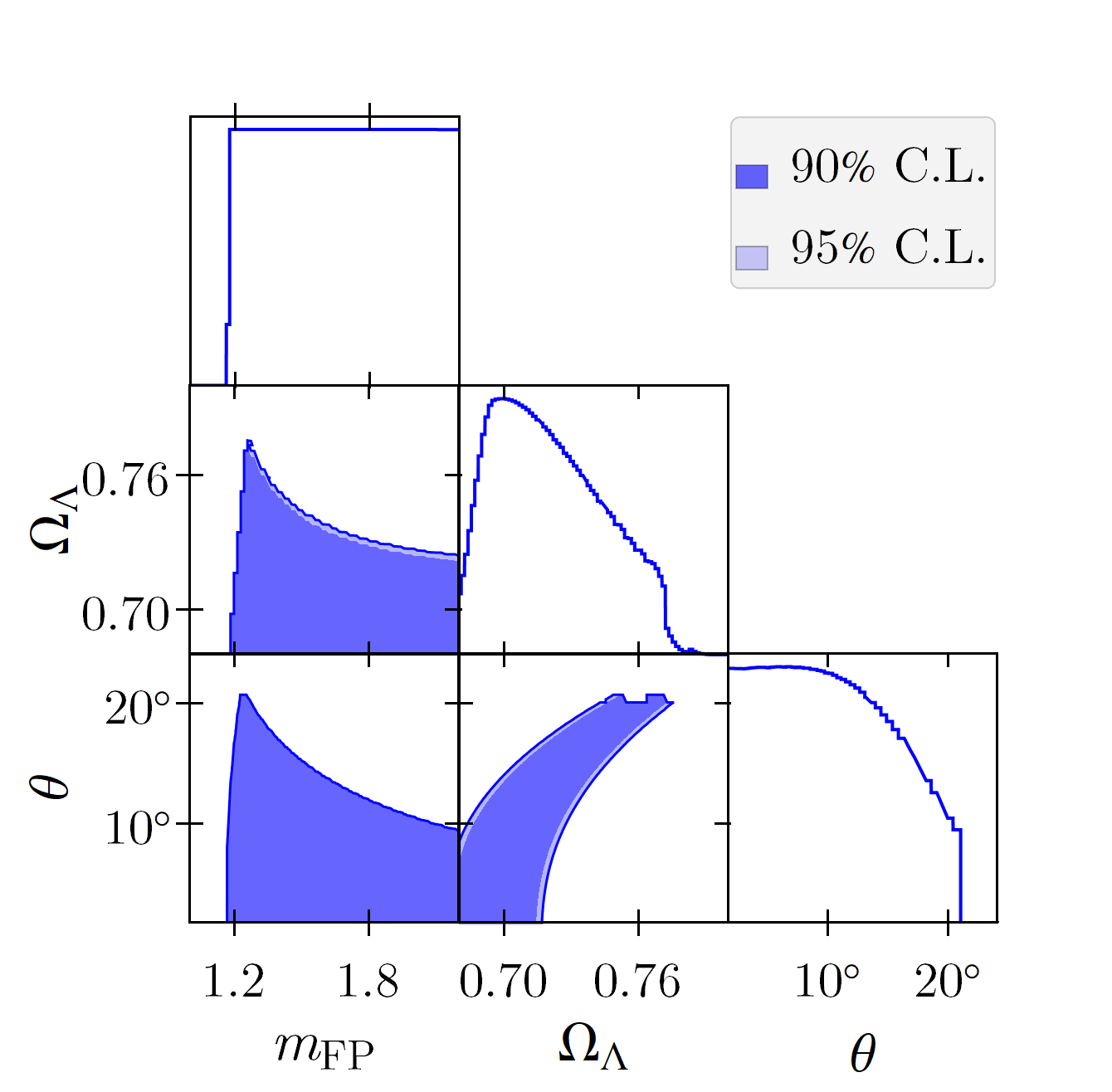}
		\caption{$B_1 B_2 B_3$ (minimal) models.}
		\label{fig:cosmotestsB123}
	\end{subfigure}
	\hfill
	\begin{subfigure}[b]{0.49\textwidth}
		\centering
		\includegraphics[width=\textwidth]{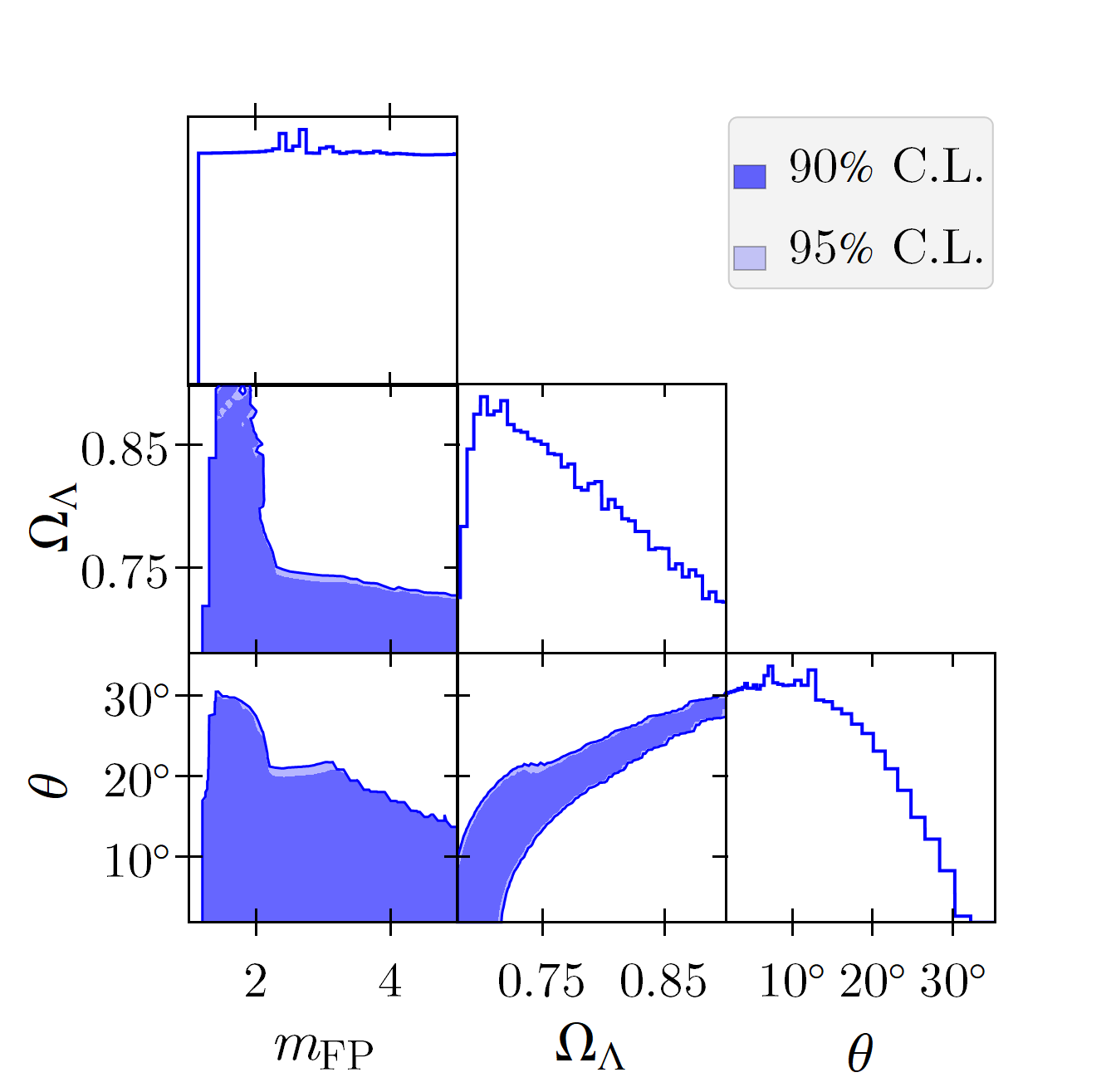}
		\caption{Self-accelerating models ($B_1 B_2 B_3 B_4$).}
		\label{fig:cosmotestsSA}
	\end{subfigure}
	\hfill
	\begin{subfigure}[b]{0.49\textwidth}
		\centering
		\includegraphics[width=\textwidth]{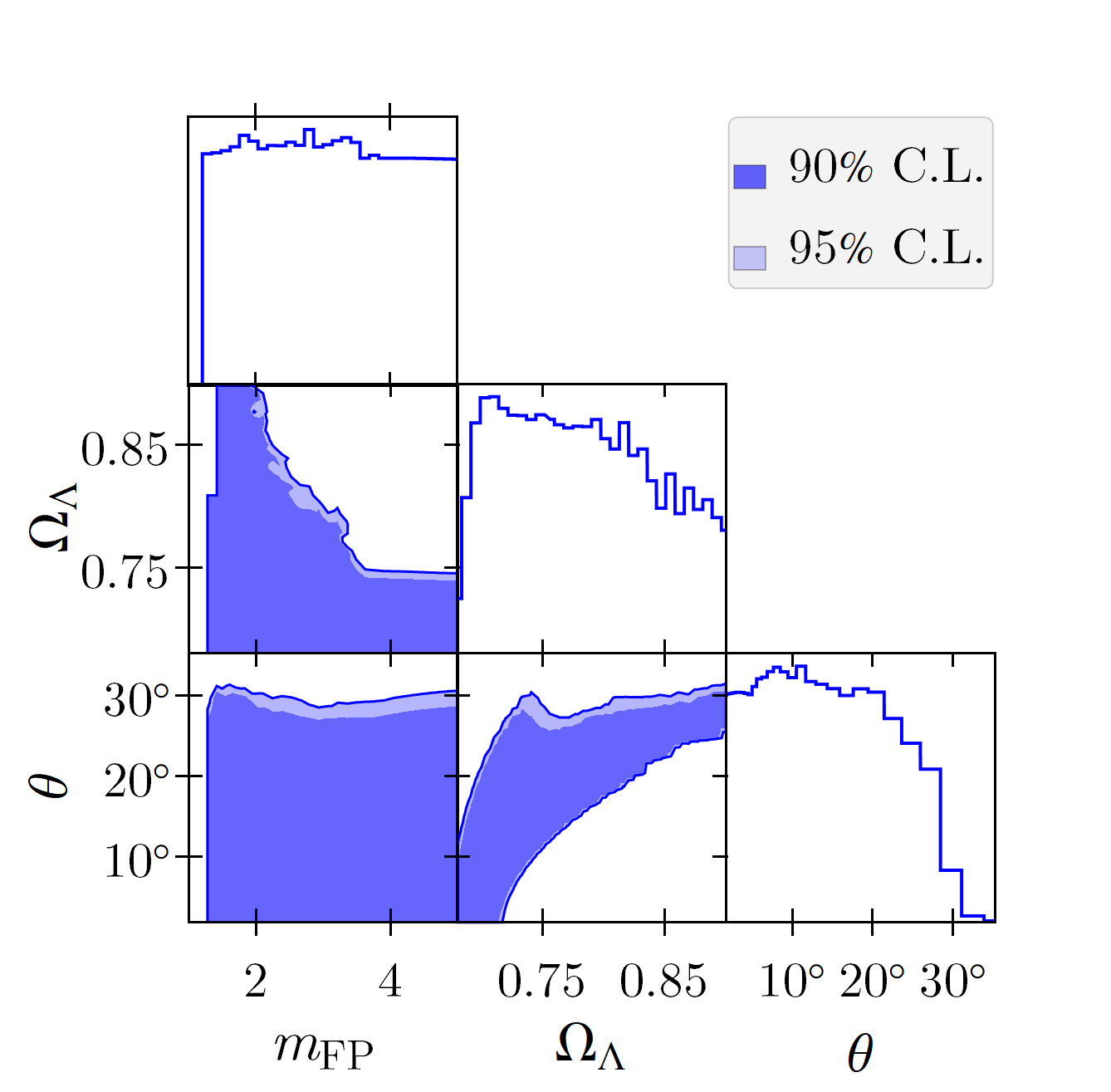}
		\caption{General models ($B_0 B_1 B_2 B_3 B_4$).}
		\label{fig:cosmotestsGen}
	\end{subfigure}
		\hfill
	\begin{subfigure}[b]{0.49\textwidth}
		\centering
		\includegraphics[width=\textwidth]{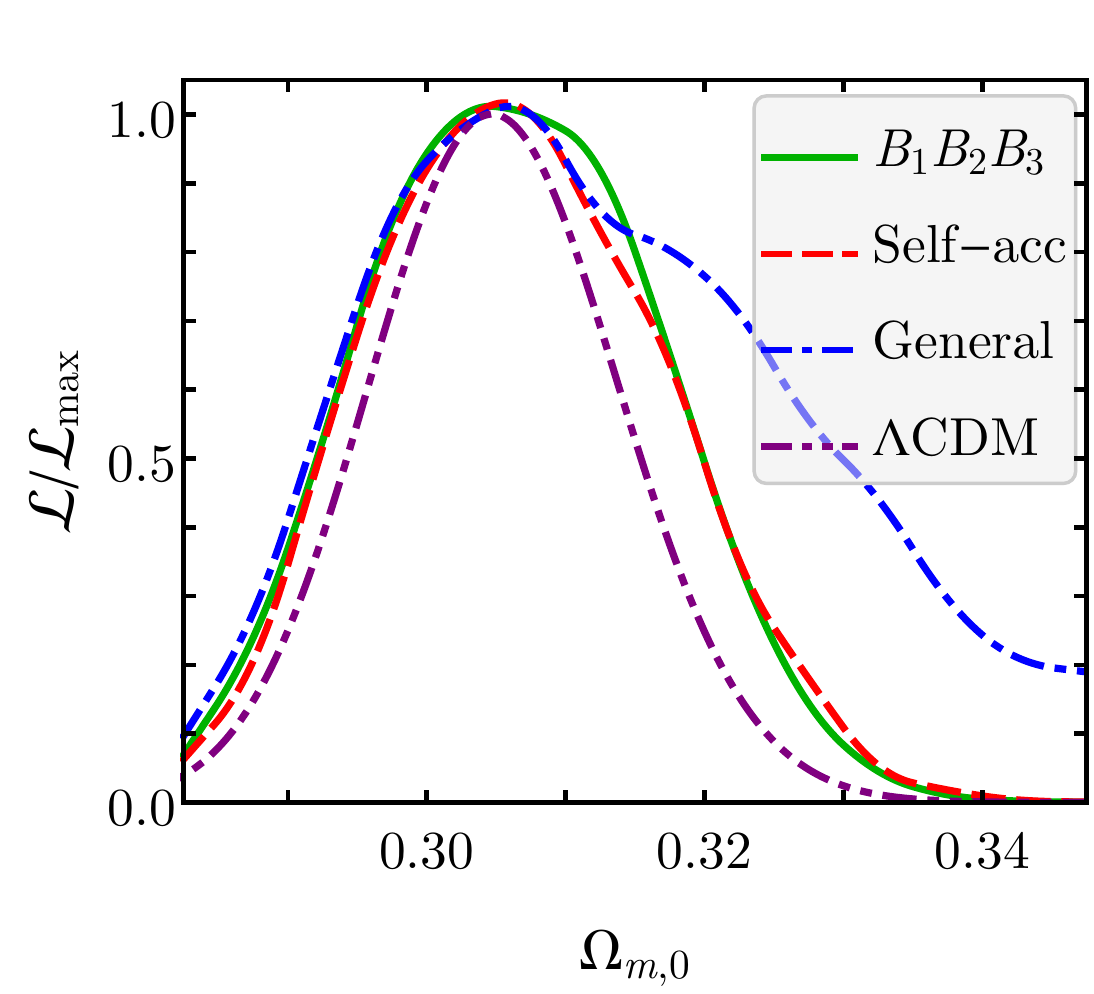}
		\caption{Matter density today, all models.}
		\label{fig:omegam0}
	\end{subfigure}
	\caption{(a)-(c): Two-dimensional (marginalized) confidence contours in the physical parameters $\theta$ (mixing angle), $\mfphat$ (graviton mass), and $\omegaleff$ (effective cosmological constant) when fitting to data from CMB/BAO+SNIa+XCL and imposing the analytical constraints of Section~\ref{sec:AnConstr}. The one-dimensional plots (on the diagonals) are normalized likelihoods, $\mathcal{L}/\mathcal{L}_\mathrm{max}$.}
	\label{fig:cosmotests}
\end{figure}

\begin{figure}[t]
	\centering
	\includegraphics[width=0.9\linewidth]{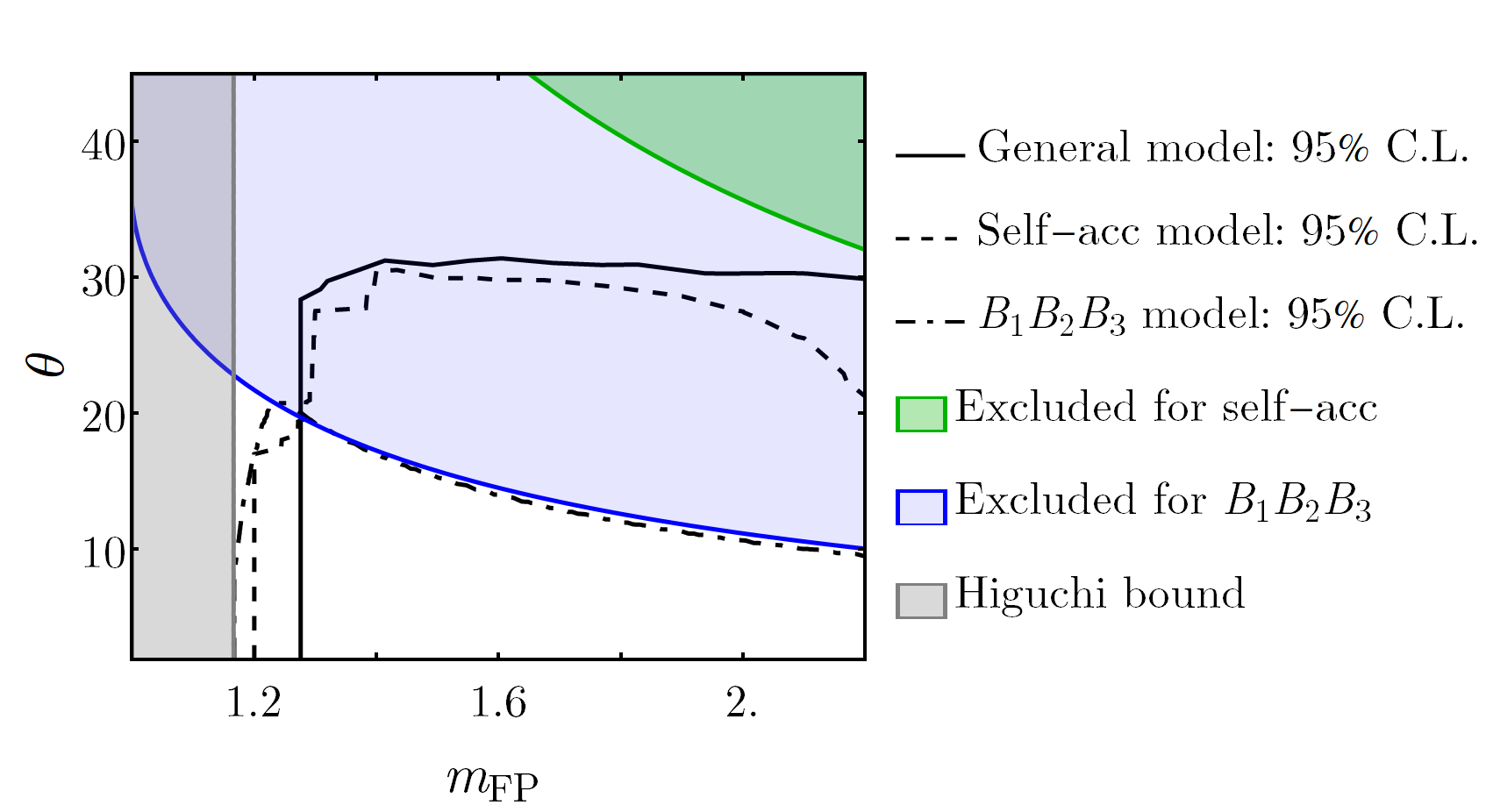}
	\caption{Marginalized $95 \, \%$ confidence contours (below the curves) for $B_1 B_2 B_3$, self-accelerating ($B_1 B_2 B_3 B_4$), and general ($B_0 B_1 B_2 B_3 B_4$) bimetric models. The colored regions are excluded due to the analytical constraints. Gray region: Higuchi bound, applies to all models. Green region: excluded region for self-accelerating solutions, as approximated by \eqref{eq:kappalimSA}. Blue region: excluded region for $B_1 B_2 B_3$ models as approximated by \eqref{eq:AnConstrB123}.}
	\label{fig:mfpthetacompiled}
\end{figure}

\paragraph{$B_1 B_2 B_3$ (minimal) models.} These models are a subset of the self-accelerating models and are of special interest since the cosmological constant terms $B_0$ and $B_4$ are absent. Still, the bimetric energy density $\omegade$ is dynamical and contributes to the accelerated expansion of the Universe. In Fig.~\ref{fig:cosmotestsB123} we show the two-dimensional marginalized confidence contours and parameter likelihoods of the $B_1 B_2 B_3$ models. The two-dimensional marginalized confidence contours are defined as level curves of $\Delta \chi^2 \equiv \chi^2 - \chi^2_\mathrm{min}$ where $90 \, \%$ corresponds to $\Delta \chi^2 = 4.61$ and $95 \, \%$ to $\Delta \chi^2 = 5.99$. To constrain the values of the physical parameters, we define the one-dimensional $90 \, \%$ confidence interval as that within which $\Delta \chi^2 < 2.71$. In Fig.~\ref{fig:mfpthetacompiled} we compile the marginalized $95 \, \%$ confidence contours in the $\mfphat \theta$-plane with the analytical constraints for general models, self-accelerating models, and $B_1 B_2 B_3$ models.

\begin{figure}[t]
	\centering
	\includegraphics[width=\linewidth]{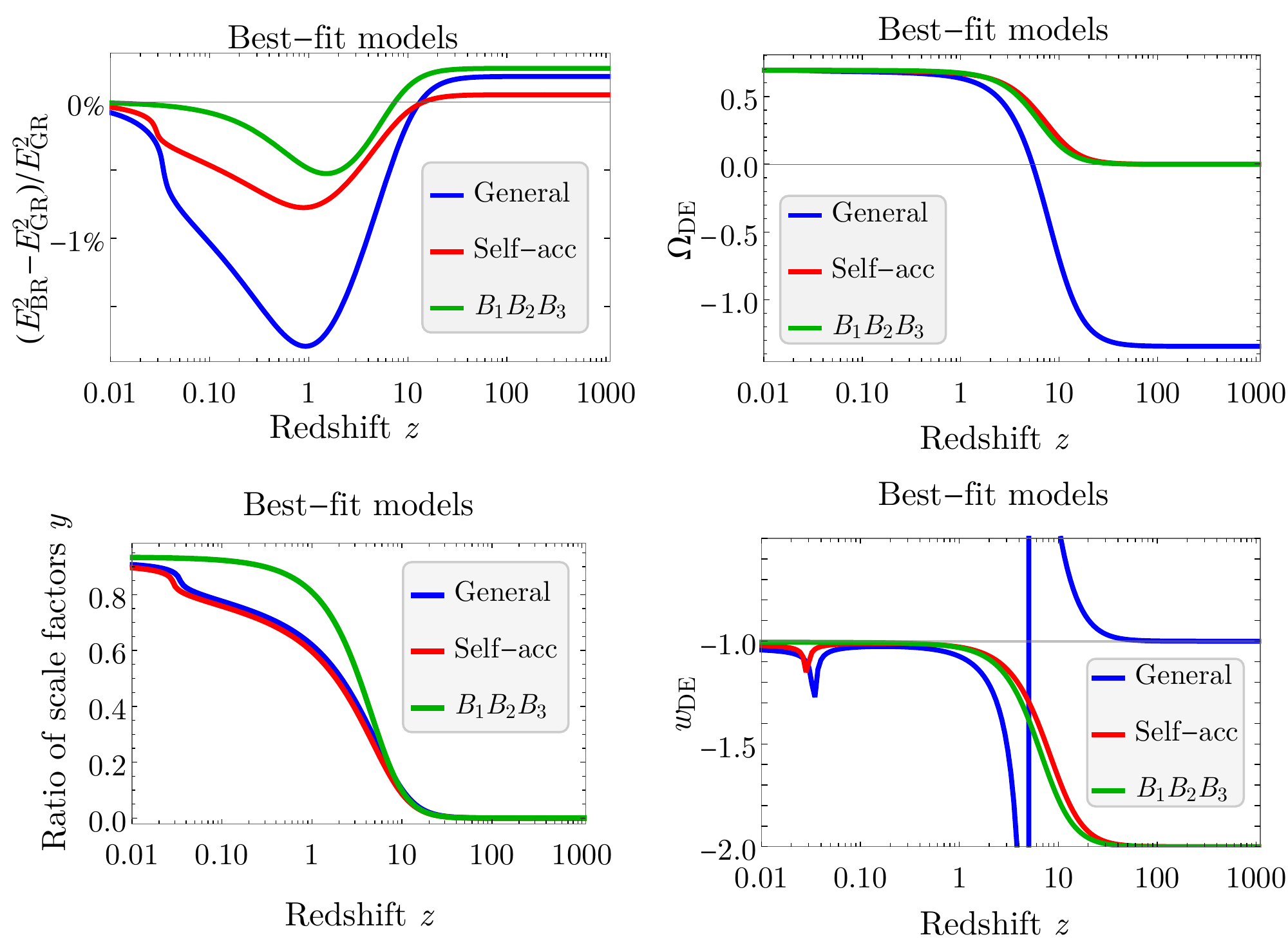}
	\caption{Best-fit models. \emph{Upper left panel:} comparing the best-fit bimetric expansion rate, $E_\mathrm{BR}$, with the best-fit $\Lambda$CDM expansion rate, $E_\mathrm{GR}$. The expansion best-fit models follow the $\Lambda$CDM model closely, except in the range $0.01 \lesssim z \lesssim 10$. \emph{Upper right panel}: the dark energy density $\omegade$. For the general and self-accelerating models, there is a steep increase in $\omegade$ around $z=0.03$ which is due to the dip in the dark energy equation of state at that redshift. \emph{Lower left panel}: ratio of the scale factors $y$. \emph{Lower right panel}: Dark energy equation of state $w_\mathrm{DE}$. There is a divergence in $w_\mathrm{DE}$ at $z \simeq 5$ for the best-fit general model, which is due to $\omegade$ crossing zero. There is no physical singularity at this point.}
	\label{fig:bestfit}
\end{figure}

The massive spin-2 field effectively provides a dynamical phantom dark energy component which is influential in the range $0 \lesssim z \lesssim 10$, see Fig.~\ref{fig:bestfit}. The minimal $\chi^2$ value is $\chi_\mathrm{min}^2 = 52.2$, hence there is a slight improvement in the fit compared with flat $\Lambda$CDM. The constraints on the physical parameters from these data sets and the analytical constraints are shown in Tab.~\ref{tab:cosmoconstr}.

The mixing angle $\theta$ has an upper bound $\theta < 21^\circ$ ($90 \, \%$ confidence) which is consistent with the approximate analytical bound $\theta \lesssim 20^\circ$ \eqref{eq:ThetaUpperBound}, see also Fig.~\ref{fig:anconstrspecmodels}. Remember that the mixing angle parameterizes the proportions of the massive and massless gravitons in the physical metric. Interestingly, there are models with a substantial mixing angle that still have viable background cosmology and local solutions (remember that our parameter choice ensures a working screening mechanism). The graviton mass has a lower limit $\mfphat \gtrsim 1.2$ ($90 \, \%$ confidence) which is due to the Higuchi bound $\mfphat^2 > 2 \omegaleff$ plus the fact that the effective cosmological constant has a lower bound, $\omegaleff = 0.70_{-0.02}^{+0.07}$ ($90 \, \%$ confidence).

There is a degeneracy between $\theta$ and $\omegaleff$, see the ``banana-like" shape in the $\omegaleff \theta$-plane of Fig.~\ref{fig:cosmotestsB123}. The dark energy density increases with time and $\omegaleff$ is the cosmological constant in the final de Sitter phase. For large values of the mixing angle $\theta$, we are far away from that phase today (i.e., at $z=0$) and hence the cosmological constant in the de Sitter phase i greater than today. The current matter density follows a distribution centered around $\omegamnot = 0.31$, see Fig.~\ref{fig:omegam0}. The best-fit parameter values and constraints are summarized in Tab.~\ref{tab:cosmoconstr} and the expansion history for the best-fit model is plotted in Fig.~\ref{fig:bestfit}. In terms of the $B$-parameters, the best-fit point is $(B_0,B_1,B_2,B_3,B_4)_\mathrm{best \; fit} \simeq (0,1.5,-1.0,0.5,0)$ and $\mathcal{O}(B_n^\mathrm{best \; fit})=1$.\\

\begin{table}[t]
	\renewcommand*{\arraystretch}{1.4}
	\centering
	\begin{tabular}{l | c | c | c | c | c | c | c | c}
		\hline\hline
		& $\theta$ & $\mfphat$ & $\omegaleff$ & $\alpha$ & $\beta$ & $\omegamnot$ & $\chi_\mathrm{min}^2$ & DoF\\
		\hline
		General model & ${11^{\circ}}^{+17^{\circ}}_{-11^\circ}$ & $2.8_{-1.6}^{+\infty}$ & $0.71_{-0.03}^{+0.25}$ & $-5.1_{-\infty}^{+\infty}$ & $34_{-32}^{+\infty}$ & $0.31^{+0.03}_{-0.02}$ & 52.0 & 5 \\
		\hline
		Self-accelerating & ${7^{\circ}}_{-7^\circ}^{+21^{\circ}}$ & $2.7_{-1.6}^{+\infty}$ & $0.70_{-0.02}^{+0.18}$ & -- & $29_{-27}^{+\infty}$ & $0.31_{-0.02}^{+0.02}$ & 52.0 & 4\\
		\hline
		$B_1 B_2 B_3$ & ${6^{\circ}}_{-6^\circ}^{+15^{\circ}}$ & $1.6_{-0.4}^{+\infty}$ & $0.70_{-0.02}^{+0.07}$ & -- & -- & $0.31_{-0.02}^{+0.02}$ & 52.2 & 3\\
		\hline
		$\Lambda$CDM & -- & -- & $0.70_{-0.01}^{+0.01} $ & -- & -- & $0.31_{-0.01}^{+0.01}$ & 52.5 & 1\\
		\hline\hline
	\end{tabular}
	\caption{Constraints on the physical parameters ($90 \, \%$ confidence) from the analytical constraints of Section~\ref{sec:AnConstr}, combined with CMB/BAO+SNIa+XCL data. The graviton mass $\mfphat$ is expressed in units of $H_0 = 100h \, \mathrm{km/s/Mpc} = 2.1h \times 10^{-33} \, \mathrm{eV}/c^2$. DoF (degrees of freedom) denotes the number of free parameters. (The upper constraint on $\omegaleff$ for the general model is calculated by linear extrapolation of the one-dimensional likelihood in the range $\omegaleff \in [0.86,0.90]$.)}
	\label{tab:cosmoconstr}
\end{table}

\paragraph{Self-accelerating models ($B_1 B_2 B_3 B_4$).} Similar results hold for the general self-accelerating models. A novel feature that appears however, is a brief phase around $z \simeq 0.03$ where the equation of state has a sharp dip, see Fig.~\ref{fig:bestfit}. Since we have one additional free parameter compared to the $B_1 B_2 B_3$ models, the marginalized confidence contours are broadened, compare Figs.~\ref{fig:cosmotestsB123} and \ref{fig:cosmotestsSA} and the constraints on the physical parameters are weaker, see Tab.~\ref{tab:cosmoconstr}. In terms of the $B$-parameters, the best-fit point is $(B_0,B_1,B_2,B_3,B_4)_\mathrm{best \; fit} \simeq (0,2.4,-2.9,3.4,-4.0)$. Thus, $\mathcal{O}(B_n^\mathrm{best \; fit})=1$. For the expansion history of the best-fit model, see Fig.~\ref{fig:bestfit}. For the two-dimensional confidence contours in the parameters $\theta$, $\mfphat$, and $\omegaleff$, see Fig.~\ref{fig:cosmotestsSA}. In Fig.~\ref{fig:sacosmofitfull} of Appendix~\ref{sec:MargLike}, we plot the confidence contours also in $\beta$. The combined constraints from the analytical constraints and cosmological data sets are summarized in Tab.~\ref{tab:cosmoconstr}.\\

\paragraph{General models ($B_0 B_1 B_2 B_3 B_4$).} For the most general bimetric models, the marginalized confidence contours are broadened compared to the self-accelerating models (compare Figs.~\ref{fig:cosmotestsSA} and \ref{fig:cosmotestsGen}) and the constraints on the physical parameters are weaker (see Tab.~\ref{tab:cosmoconstr}). In terms of the $B$-parameters, the best-fit point is $(B_0,B_1,B_2,B_3,B_4)_\mathrm{best \; fit} \simeq (-4.0,7.0,-8.2,9.7,-11.3)$, that is, $\mathcal{O}(B_n^\mathrm{best \; fit}) = 10$. The expansion history for the best-fit model is plotted in Fig.~\ref{fig:bestfit} and the two-dimensional confidence contours of $\theta$, $\mfphat$, and $\omegaleff$ are plotted in Fig.~\ref{fig:cosmotestsGen}. The confidence contours in the parameters $\alpha$ and $\beta$ are plotted in Fig.~\ref{fig:gencosmofitfull} in Appendix~\ref{sec:MargLike}. The combined constraints on the physical parameters from the analytical constraints and cosmological data sets are summarized in Tab.~\ref{tab:cosmoconstr}. In Fig.~\ref{fig:mfpthetacompiled} we compile the marginalized $95 \, \%$ confidence contours in the $\mfphat \theta$-plane with the analytical constraints for general models, self-accelerating models, and $B_1 B_2 B_3$ models.

Interestingly, the best-fit $\chi^2$ value does not improve when going from the self-accelerating models to the general bimetric models, see Tab.~\ref{tab:cosmoconstr}. With respect to the data sets employed here, one may conclude that the self-accelerating models are preferred compared to the general bimetric models due to the smaller number of free parameters. In other words, the best-fit self-accelerating model is, to good approximation, the best-fit cosmological model.\\

\noindent The results we have presented are sensitive the data sets employed. For example, if we only consider statistical errors in the SNIa data, there is a more pronounced peak in the likelihood of the physical parameters and the best-fit is far away from any GR limit. Increasing the ratio of the sound horizons at the drag epoch ($z_d \simeq 1060$) and photon decoupling ($z_* \simeq 1090$), the likelihood becomes even more pronounced and the values of all the physical parameters can be constrained with $90 \, \%$ confidence, see Appendix~\ref{sec:MargLike} for an example. On the other hand, if we add the shift parameter $\mathcal{R}$ at this level, the peak in the likelihood is still far away from any GR limit but is less pronounced than without $\mathcal{R}$. To summarize, the observational constraints on the physical parameters depend on what data sets that we use. Here, we have followed a very conservative approach and used the CMB/BAO ratio, excluding the $\mathcal{R}$ parameter (see Appendix~\ref{sec:DataSets} for more details).\\

\paragraph{Information criteria.} As evident from Tab.~\ref{tab:cosmoconstr}, bimetric cosmology and $\Lambda$CDM provide, to close approximation, equally good fits to our data sets whereas the former has a larger number of parameters to accomplish this. As a rough estimate of how well bimetric cosmology performs compared to $\Lambda$CDM, we compute the Akaike information criterion (AIC),
\begin{equation}
\mathrm{AIC} \equiv 2N_\mathrm{param} + \chi^2_\mathrm{min},
\end{equation}
and Bayesian information criterion (BIC),
\begin{equation}
\mathrm{BIC} = N_\mathrm{param} \ln N_\mathrm{data} + \chi^2_\mathrm{min},
\end{equation}
where $N_\mathrm{param}$ is the number of parameters in the model and $N_\mathrm{data}$ is the number of data points \cite{Liddle:2007fy}. The results are presented in Tab.~\ref{tab:ICs}. With a larger number of parameters, the performance, as assessed by the information criteria, is worse for the bimetric models, since flat $\Lambda$CDM provides such a good fit to these data sets. However, the theory may have beneficial properties that are not quantified by the information criteria, such as self-accelerating solutions.

\begin{table}[t]
	\centering
	\begin{tabular}{c | c c c c}
		\hline\hline
		& $\Lambda$CDM (flat)& General bimetric & Self-accelerating & $B_1 B_2 B_3$\\
		\hline
		AIC & 54.5 & 62.0 & 60.0 & 58.2\\
		BIC & 56.4 & 71.7 & 67.7 & 64.0\\ \hline
		$\Delta$AIC & 0 & $+7.5$ & $+5.5$ & $+3.7$\\
		$\Delta$BIC & 0 & $+15.3$ & $+11.3$ & $+7.6$\\
		\hline\hline
	\end{tabular}
	\caption{Information criteria. Here, we have set $\Lambda$CDM as our reference model so that $\Delta \mathrm{IC} = \mathrm{IC}_\mathrm{model} - \mathrm{IC}_\mathrm{\Lambda CDM}$. A $\Delta \mathrm{AIC} > +5$ is commonly regarded as strong preference for the reference (flat $\Lambda$CDM) model compared to the (bimetric) model \cite{Liddle:2007fy}.}
	\label{tab:ICs}
\end{table}

To summarize, bimetric cosmology and flat $\Lambda$CDM both provide good fits to our cosmological data sets, with similar $\chi_\mathrm{min}^2$. Hence, the bimetric physical parameters are unconstrained, except the effective cosmological constant and the upper bound on $\theta$. At this level, the one-dimensional likelihoods of $\mfphat$, $\alpha$, and $\beta$ are approximately flat and the only constraints on them are the analytical ones, see Section~\ref{sec:AnConstr}. Thanks to a working screening mechanism, the best-fit cosmological parameters are compatible with observations from solar system tests, strong lensing by galaxies, and galaxy cluster lensing (see the previous subsections). Constraints from gravitational waves, which constrain $\theta \lesssim 20^\circ$ in the graviton mass range $10^{11} \lesssim \mfphat \lesssim 10^{12}$ (see Fig.~\ref{fig:localtestscompiled}), are also satisfied. Due to the presence of four new parameters, bimetric cosmology is a more flexible model than $\Lambda$CDM. Interestingly, since there are viable bimetric models also away from the GR limits, there is a possibility that other data sets can break the draw between the two. In particular, taking into account local measurements of the Hubble constant introduces a discrepancy in the $\Lambda$CDM model. With a more flexible background evolution, bimetric cosmology might be able to alleviate the tension. Also, with a framework for analyzing cosmological perturbations/structure formation, it may be possible to put more stringent constraints on the physical parameters.

\subsection{The Hubble tension}
Let us assess very roughly how well we expect the best-fit bimetric models to perform in relation to the Hubble tension. Since the expansion is close to the best-fit $\Lambda$CDM model for redshifts $z \gtrsim 10$ (see Fig.~\ref{fig:bestfit}), we assume here that the sound speed at photon decoupling takes the same value as for a $\Lambda$CDM model. Because the expansion rate of the best-fit bimetric model is smaller in the range $0.01 \lesssim z \lesssim 10$, the integral $I_* = \int_0^{z_*} dz/E(z)$ is greater than the $\Lambda$CDM case. Therefore, in order for $\ell_A$ to stay the same, $H_0$ must be greater, see \eqref{eq:lAdef}. Calculating the integral for the best-fit bimetric and $\Lambda$CDM models, we get $H_0 = 1.004 \, H_0^\mathrm{\Lambda CDM}$ for the general model and $H_0 = 1.002 \, H_0^\mathrm{\Lambda CDM}$ for the self-accelerating model. Thus, instead of the $\Lambda$CDM value $H_0^\mathrm{\Lambda CDM} = 67.9 \, \mathrm{km/s/Mpc}$ \cite{Ade:2015xua} we get a value which is greater by $0.4 \, \%$ for the general model, $H_0 = 68.2 \, \mathrm{km/s/Mpc}$, and a value that is greater by $0.2 \, \%$, $H_0 = 68.0 \, \mathrm{km/s/Mpc}$, for the self-accelerating model, which is somewhat closer to the value from the late-time probes $\simeq 73 \, \mathrm{km/s/Mpc}$ \cite{Riess:2016jrr}. For the best-fit $B_1 B_2 B_3$ model, the Hubble constant is to good approximation unchanged compared with the $\Lambda$CDM value. Due to the presence of four new parameters, bimetric cosmology is a more flexible model than $\Lambda$CDM and it is possible that there are viable models which increase the value of $H_0$ significantly. Ref. \cite{Mortsell:2018mfj} concluded that bimetric two-parameter models can increase the Hubble constant somewhat but not completely alleviate the tension. However, such models are restrictive and do not comply with the constraints from demanding a working screening mechanism and a consistent cosmology. In Ref. \cite{Lindner:2020eez}, the Hubble tension is studied for general bimetric models. However, they use the shift parameter $\mathcal{R}$ as a data point, which introduces a number of assumptions, see Appendix~\ref{sec:DataSets} for a detailed discussion. Whether or not the tension can be alleviated is a question for future study.

\section{Conclusions and outlook}
\label{sec:ConclOut}
We have studied the observational constraints on the physical parameters of bimetric gravity using CMB/BAO, SNIa, and a background independent measurement of $\omegamnot$ (XCL) to fit for the background cosmology. Using the analytical constraints of \cite{PhysParamTh}, we ensure a real-valued background cosmology devoid of the Higuchi ghost and a working Vainshtein screening mechanism for the local solutions. These constraints can be expressed analytically in terms of the physical parameters. Thereby, we ensure that gravitational tests ranging from solar system scales to galaxy cluster scales, are satisfied. (Constraints from gravitational wave observation are also compatible with our results.) The results are summarized in Figs.~\ref{fig:cosmotests} and \ref{fig:mfpthetacompiled} and Tab.~\ref{tab:cosmoconstr}. For a general bimetric model, there is an upper limit on the mixing angle, $\theta < 28^\circ$ ($90 \, \%$ credibility). Hence, the physical metric can contain a significant amount of the massive spin-2 field. There is a lower limit on the graviton mass, $\mfphat > 1.2$ ($90 \, \%$ credibility). In units of $\mathrm{eV}/c^2$, $\mfphat > 2.5h \times 10^{-33} \, \mathrm{eV}/c^2$ corresponding to a Compton wavelength of the size of the observable universe. The lower bound is many orders of magnitude smaller than the mass of the standard model particles. The effective cosmological constant, that is the dark energy density in the asymptotic future, is $\omegaleff = 0.71_{-0.03}^{+0.25}$ ($90 \, \%$ credibility) and the current matter density $\omegamnot = 0.31_{-0.02}^{+0.03}$ ($90 \, \%$ credibility). Interestingly, the general bimetric models (five free parameters) and the self-accelerating models (four free parameters) provide equally good fits, suggesting that the latter models are preferred over the former by these data sets. In the self-accelerating models, the accelerated expansion is due to the dynamical massive spin-2 field, without a cosmological constant

\noindent Since the bimetric models improve the $\chi^2_\mathrm{min}$ only somewhat, compared to the flat $\Lambda$CDM model, the likelihoods of the physical parameters $\mfphat$, $\alpha$, and $\beta$ are approximately flat and hence there are no bounds on these parameters, except the analytical constraints. However, the results depend on the data sets employed. For example, if we set a greater value of the ratio of the sound horizons at the drag epoch and photon decoupling ($z_d \simeq 1060$ and $z_* \simeq 1090$, respectively), the improvement in the fit can be substantial and the physical parameters constrained by cosmology. Here, we have used the most recent data sets and combined CMB and BAO in a way that effectively cancels the dependence on the cosmology before photon decoupling.

The fact that there are viable models even for a relatively large mixing angle ($\theta \lesssim 28^\circ$) opens the possibility to break the draw between bimetric cosmology and $\Lambda$CDM by introducing new data sets. In particular, taking late-time measurements of the Hubble constant into account, there appears a discrepancy in the $\Lambda$CDM model. A back-of-the-envelope calculation indicates that the best-fit bimetric model increases the value of the Hubble constant by $\simeq 0.4 \, \%$, easing the tension to a small degree. Whether or not it can be alleviated completely is a subject for future research.

\acknowledgments
Thanks to Angelo Caravano and Marvin Lüben for many interesting discussions on the subject and to Francesco Torsello for comments on the manuscript. This research utilized the Sunrise HPC facility supported by the Technical Division at the Department of Physics, Stockholm University. Special thanks to Mikica Kocic for technical support.\\

\noindent \textbf{Note added:} Ref.~\cite{Caravano:2021aum} appeared at the same time as this paper. In the former paper, the authors analyze the constraints on the physical parameters from cosmological data and local observations for bimetric models with three parameters or less. Their results are consistent with ours, when comparable. \newpage

\appendix
\section{Marginalized likelihoods}
\label{sec:MargLike}
\begin{figure}[t]
	\centering
	\includegraphics[width=0.7\linewidth]{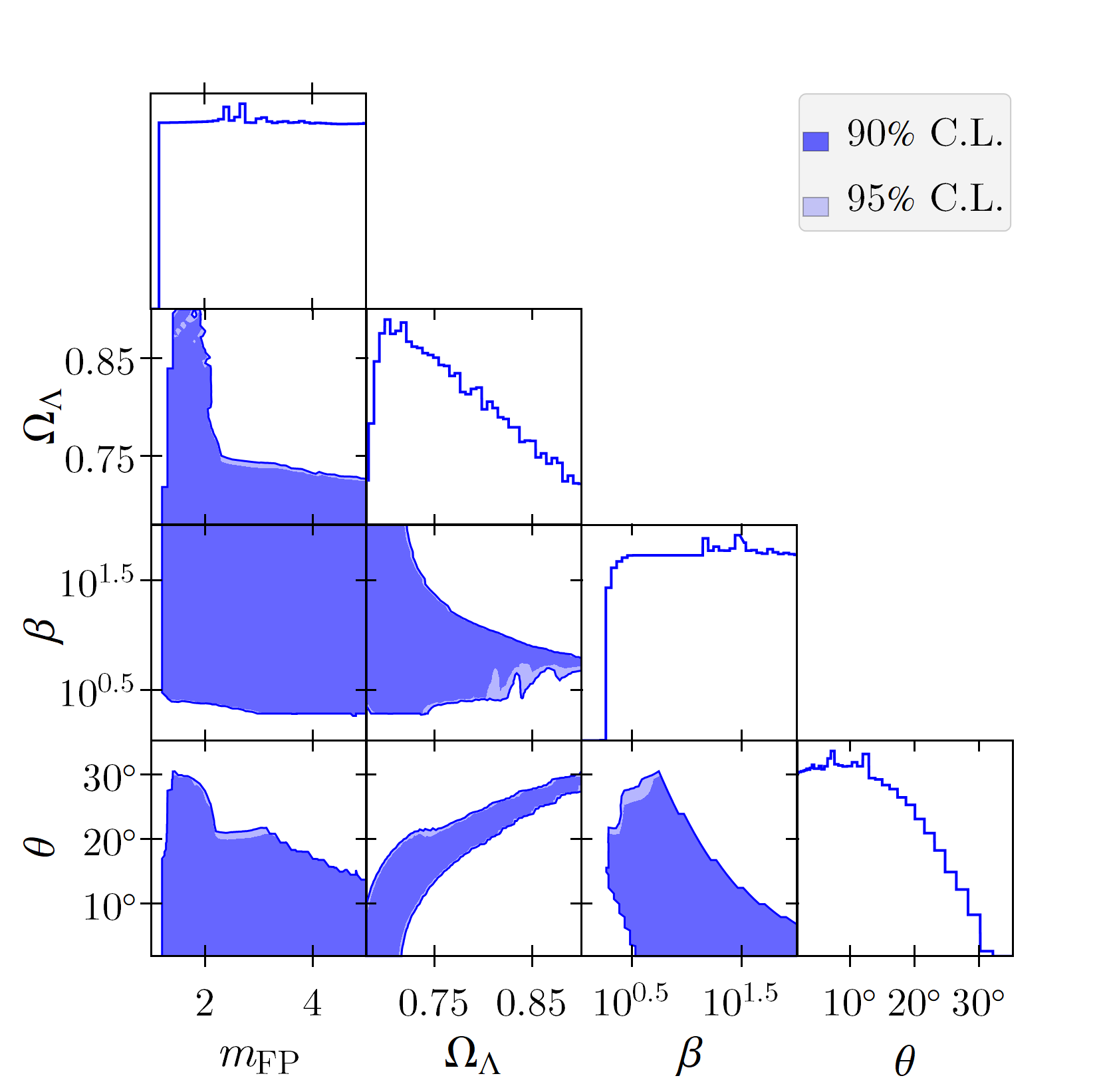}
	\caption{Confidence contours and normalized likelihoods for the self-accelerating models. As expected, the likelihoods of $\mfphat$ and $\beta$ are approximately flat above the thresholds set by the analytical constraints, $\mfphat \gtrsim 1$ and $\beta \gtrsim 1$. The former is due to the Higuchi bound, $\mfphat^2 > 2 \omegaleff$, and the latter is due to the requirement of a working Vainshtein screening mechanism. Note that $\mfphat$ and $\beta$ behave similarly.}
	\label{fig:sacosmofitfull}
\end{figure}

\begin{figure}
	\centering
	\includegraphics[width=0.9\linewidth]{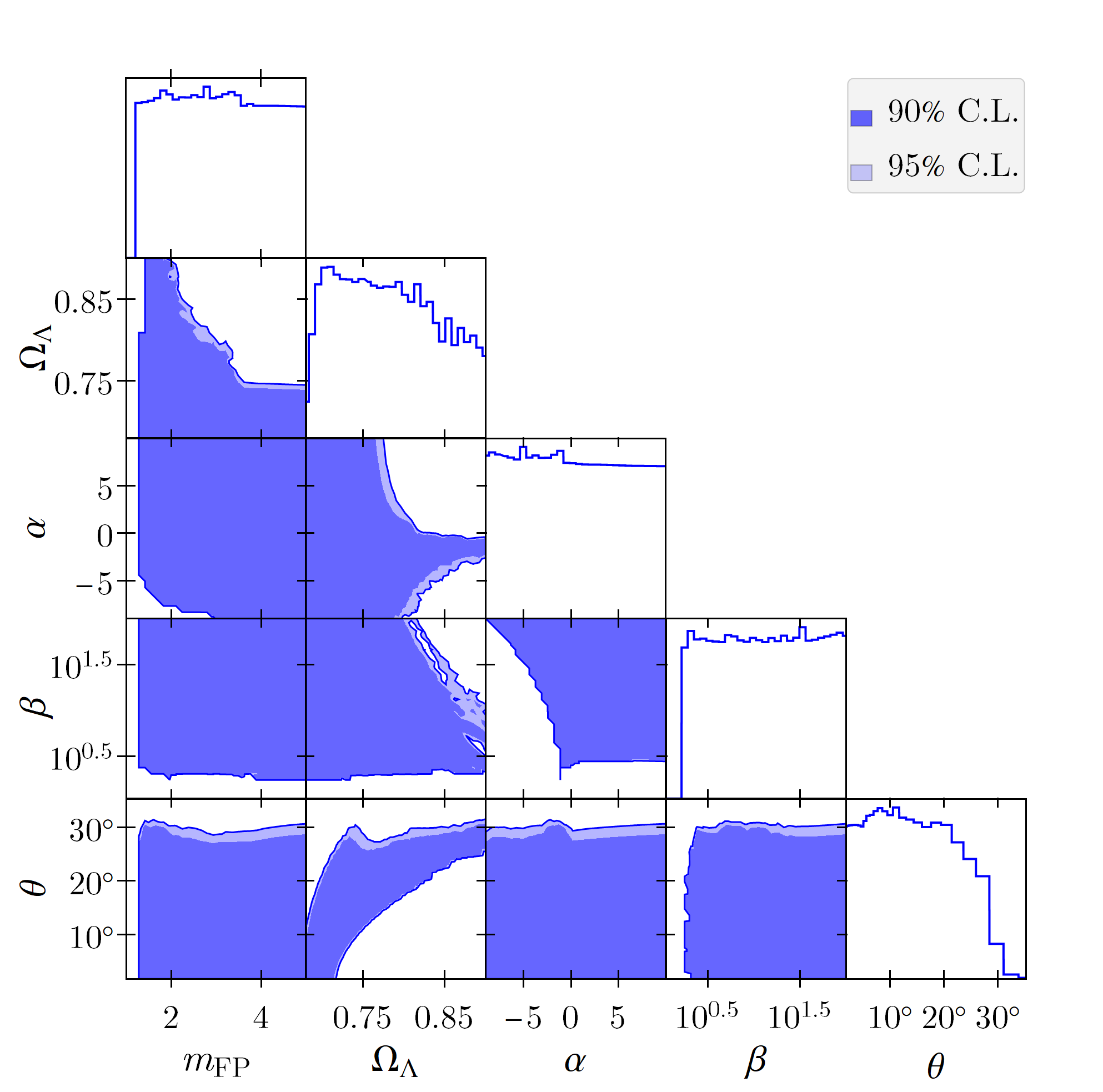}
	\caption{Confidence contours and normalized likelihoods for the general bimetric models. As expected, the likelihood of $\alpha$ is flat and the likelihoods of $\mfphat$ and $\beta$ are approximately flat above the thresholds set by the analytical constraints, $\mfphat \gtrsim 1$ and $\beta \gtrsim 1$. Thus, $\mfphat$, $\alpha$, and $\beta$ are poorly constrained by these cosmological data sets. There is a region in the $\alpha \beta$-plane which is excluded due to the constraints from having a working screening mechanism, cf. Fig.~\ref{fig:dynhig2}.}
	\label{fig:gencosmofitfull}
\end{figure}

Here, we show the confidence contours and likelihoods in the parameters $(\theta,\mfphat,\omegaleff,\beta)$ for the self-accelerating models in Fig.~\ref{fig:sacosmofitfull} and in the parameters $(\theta,\mfphat,\omegaleff,\alpha,\beta)$ for the general models in Fig.~\ref{fig:gencosmofitfull}.

\begin{figure}
	\centering
	\includegraphics[width=0.6\linewidth]{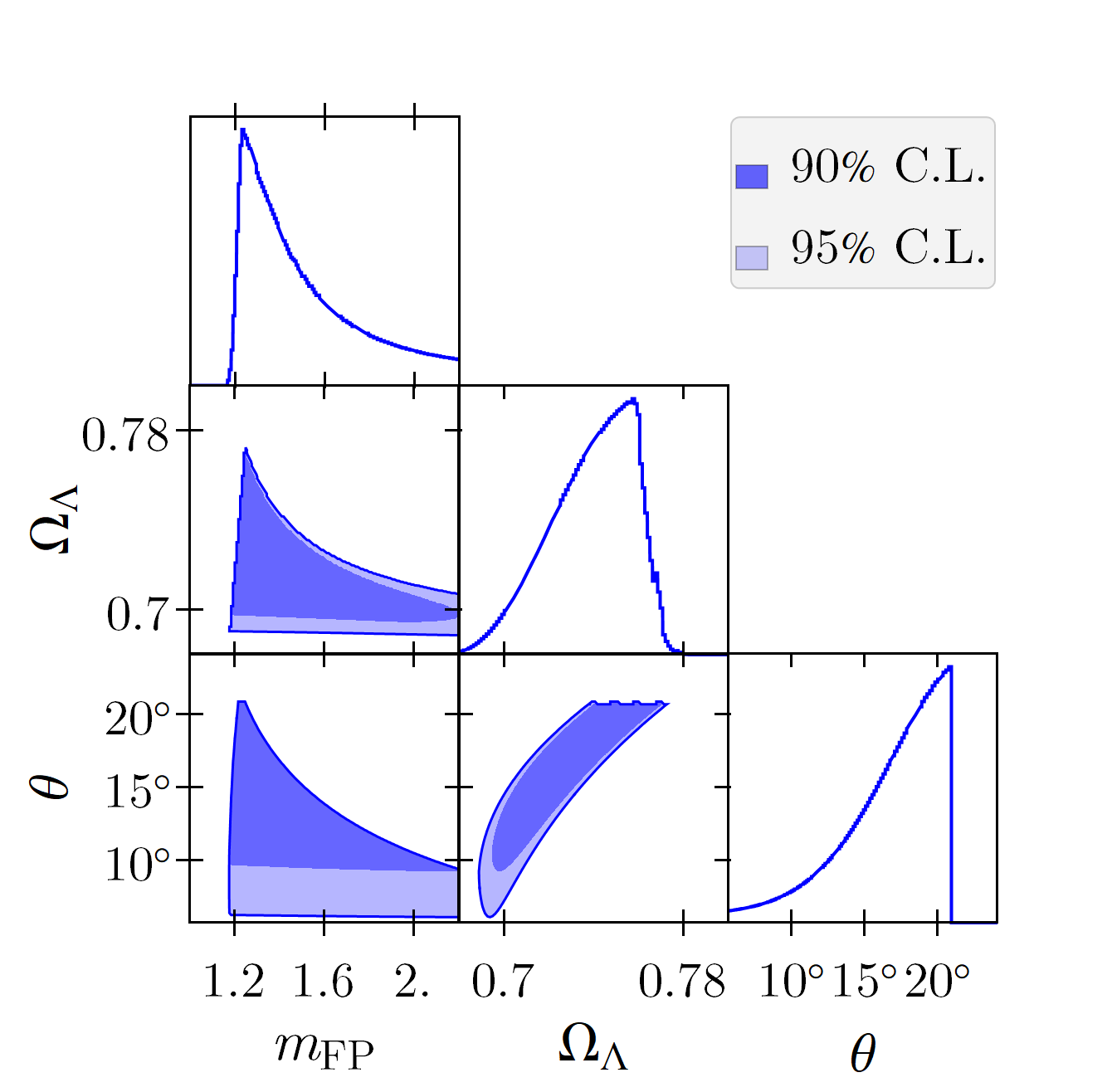}
	\caption{Confidence contours and normalized likelihoods for $B_1 B_2 B_3$ models with a different data set than in the main text. Here, $\ell_A$ is from \cite{Ade:2015rim} and the BAO points from \cite{Ade:2015xua}. The SNIa data is from Pantheon but without systematic errors and the ratio of the sound horizons is set to $r_s(z_d) / r_s(z_*) = 1.03$.}
	\label{fig:b123cosmofitalt}
\end{figure}

In Fig.~\ref{fig:b123cosmofitalt}, we show the result of fitting $B_1 B_2 B_3$ models to another data set. Here, we use CMB/BAO data with $\ell_A = 301.63 \pm 0.15$ \cite{Ade:2015rim} and the same four BAO points as in Ref. \cite{Ade:2015xua}. The SNIa data points are still from Pantheon but without systematic errors. The ratio of the sound horizons at the drag epoch and at photon decoupling is set to $r_s(z_d) / r_s(z_*) = 1.03$. As evident from Fig.~\ref{fig:b123cosmofitalt}, here the bimetric model improves the fit substantially compared with the flat $\Lambda$CDM model and the likelihood has a pronounced peak at $(\theta,\mfphat,\omegaleff) = (21^\circ,1.2,0.76)$.\newpage

\section{Data sets}
\label{sec:DataSets}
\paragraph{Background independent $\omegamnot$.} The Chandra X-ray observatory measures the X-ray gas fraction in galaxy clusters and thereby obtains a measurement of the ratio of the baryonic matter density and the total (pressureless) matter density. Combined with constraints on the physical baryonic density from big bang nucleosynthesis (BBN) $\Omega_b h^2 = 0.0214 \pm 0.0020$ and $h = 0.72 \pm 0.08$, the total matter density is $\omegamnot = 0.24 \pm 0.04$ \cite{Allen:2004cd}. Instead of the constraint on $\Omega_b h^2$ from BBN, we use a constraint from extragalactic dispersion of fast radio bursts, which is independent of the background cosmology \cite{Macquart2020}. This gives a somewhat greater value of $\Omega_b$ and hence the total matter density and the error also increases,
\begin{equation}
\omegamnot = 0.29 \pm 0.09.
\end{equation}
We refer to this data point as XCL (X-ray galaxy cluster).

\paragraph{CMB data.} It is common to represent the information contained in the CMB power spectrum by a few parameters, including the shift parameters $\ell_A$ and $\mathcal{R}$ (see e.g. \cite{Kosowsky:2002zt,Wang:2007mza}). The former is the angular scale of the sound horizon at photon decoupling, $\ell_A$, is given by (see e.g. \cite{Ade:2015rim}),
\begin{equation}
\label{eq:lAdef}
\ell_A = \pi D_{A}(z_*) / r_{s}(z_*), 
\end{equation}
where $D_A(z_*)$ is the comoving angular diameter distance to the last scattering surface $z_*$,
\begin{equation}
\label{eq:DAeq}
D_A(z) = I(z) / H_0,
\end{equation}
with $I(z)$ being,
\begin{equation}
I(z) \equiv \int_{0}^{z} \frac{dz'}{E(z')}, \quad \Omega_k=0,
\end{equation}
and $r_{s}(z_*)$ is the sound horizon at photon decoupling $z_*$. We set $z_* \simeq 1090$, using the relations in Ref.~\cite{Hu:1995en} together with the best-fit values of the physical baryon and matter densities from CMB and BAO data \cite{Aghanim:2018eyx}. The second shift parameter $\mathcal{R}$ represents the angular scale of the Hubble horizon at $z_*$ \cite{Efstathiou:1998xx}, and is commonly given by,
\begin{equation}
\label{eq:Req}
\mathcal{R} = \sqrt{\omegamnot H_0^2} D_A(z_*).
\end{equation}
For a $w$CDM model, with the 2018 Planck data, the shift parameters are constrained to be \cite{Chen:2018dbv},
\begin{equation}
\label{eq:ShiftParams}
\ell_A = 301.462^{+0.089}_{-0.090}, \quad \mathcal{R} = 1.7493 ^{+0.0046}_{-0.0047},
\end{equation}
with normalized covariance matrix,
\begin{equation}
\widehat{C} = \left(\begin{array}{cc}
1 & 0.47\\
0.47 & 1
\end{array}\right).
\end{equation}
The covariance matrix is given by $C_{ij} = \sigma_i \sigma_j \widehat{C}_{ij}$ (no summation over the indices). In some papers on bimetric cosmology, the CMB data is represented by $\ell_A$ and $\mathcal{R}$, see e.g. \cite{Dhawan:2017leu,Mortsell:2018mfj,Lindner:2020eez}. To use $\mathcal{R}$, one must introduce a number of assumptions. First, eq. \eqref{eq:Req} is applicable only if all contributions to $E(z_*)$ are negligible, except the matter density $\Omega_m(z_*)$ \cite{Komatsu:2008hk}. This is certainly not the case if $\mfphat \to \infty$, $\alpha \to \infty$, or $\beta \to \infty$. In these limits, the bimetric dark energy fluid mimics a cold dark matter component \cite{PhysParamTh} and hence we have a ``dark degeneracy" \cite{Kunz:2007rk}. In these cases, even if the background expansion history coincides with a $\Lambda$CDM concordance model,
\begin{equation}
\mathcal{R}|_{\mfphat,\alpha,\beta \to \infty} = \sqrt{(1-\omegaleff) H_0^2} D_A(z_*) / \cos \theta.
\end{equation}
Due to the $\cos \theta$ in the denominator, $\mathcal{R}$ will be far away from the tabulated value even if the background expansion follows the best-fit $\Lambda$CDM model. Second, the shift parameters are not direct observables, but are derived from the CMB power spectrum assuming a cosmological model (typically $\Lambda$CDM or $w$CDM), thus introducing assumptions of the primordial power spectrum and the growth of structure \cite{Elgaroy:2007bv}. In fact, $\mathcal{R}$ is more sensitive to these assumptions than $\ell_A$ \cite{Elgaroy:2007bv}. Since there is no established framework for treating structure formation in bimetric theory, we are conservative and combine $\ell_A$ with BAO data in a way which effectively eliminates the dependence on the cosmology before $z_* \simeq 1090$ (see below).

\paragraph{BAO data.} Baryon acoustic oscillations measures the ratio of the sound horizon scale $r_{s}(z_d)$ at the drag epoch $z_d$ and a particular cosmological distance scale, here $D_A$, $d_A$, or $D_V$ (depending on the data set), collectively denoted by $D_X$. $D_A$ is the comoving angular diameter distance \eqref{eq:DAeq}, $d_A$ is the angular diameter distance,
\begin{equation}
d_A(z) = D_A(z) /(1+z),
\end{equation}
and $D_V$ is the volume average distance,
\begin{equation}
D_V(z) = \left[D_A^2(z) z / H(z)\right]^{1/3}.
\end{equation}
We use the data sets from 6dFGS \cite{Beutler_2011}, SDSS MGS \cite{Ross:2014qpa}, BOSS DR12 \cite{Alam:2016hwk}, BOSS DR14 \cite{Bautista:2017wwp}, and eBOSS QSO \cite{Zhao:2018gvb}, in total ten points in the redshift range $z \in [0.106,1.944]$, see Tab.~\ref{tab:baodata}.

\begin{table}[t]
	\renewcommand*{\arraystretch}{1.4}
	\centering
	\begin{tabular}{c | c | c | c c c c }
		\hline\hline
		Data set & $z_\mathrm{eff}$ & Distance measure & \multicolumn{4}{c}{$10^4 \widehat{C}_{ij}$} \\
		\hline \hline
		6dFGS \cite{Beutler_2011} & $0.106$ & $D_V / r_d = 2.976 \pm 0.133$ & \multicolumn{4}{c}{---}\\
		\hline
		SDSS MGS \cite{Ross:2014qpa} & $0.15$ & $D_V / r_d = 4.466 \pm 0.168$ & \multicolumn{4}{c}{---}\\
		\hline
		& 0.38 & $D_A / r_d = 10.27 \pm 0.15$ & $10^4$ & $4970$ & $1991$ & \\
		BOSS DR12 \cite{Alam:2016hwk} & $0.51$ & $D_A /r_d = 13.38 \pm 0.18$ & $4970$ & $10^4$ & $984$ & \\
		& $0.61$ & $D_A /r_d = 15.45 \pm 0.22$ & $1991$ & $984$ & $10^4$ & \\
		\hline
		BOSS DR14 \cite{Bautista:2017wwp} & $0.72$ & $D_V / r_d = 16.08 \pm 0.41$ & \multicolumn{4}{c}{---}\\
		\hline
		& $0.978$ & $d_A / r_d = 10.7 \pm 1.9$ & $10^4$ & $4656$ & $2662$ & $248$\\
		eBOSS QSO \cite{Zhao:2018gvb} & $1.23$ & $d_A / r_d = 12.0 \pm 1.1$ & $4656$ & $10^4$ & $6130$ & $954$\\
		& $1.526$ & $d_A / r_d = 11.97 \pm 0.65$ & $2662$ & $6130$ & $10^4$ & $4257$\\
		& $1.944$ & $d_A / r_d = 12.23 \pm 0.99$ & $248$ & $954$ & $4257$ & $10^4$\\
		\hline\hline
	\end{tabular}
	\caption{BAO data sets used in our analysis. Here, $r_d = r_s(z_d)$ and the covariance matrix is given by $C_{ij} = \sigma_i \sigma_j \widehat{C}_{ij}$ (no summation implied).}
	\label{tab:baodata}
\end{table}

Here, we will be conservative and use CMB and BAO data in a combination that effectively eliminates the dependence on the cosmology before the time of photon decoupling \cite{Sollerman:2009yu}, namely,
\begin{equation}
\label{eq:CMBBAO}
\Pi_i = \pi \frac{D_X(z_i)/r_s(z_d)}{\ell_A} \frac{r_{s}(z_d)}{r_{s}(z_*)} = \frac{D_X(z_i)}{D_A(z_*)},
\end{equation}
where $i$ runs over the BAO points $z_i \in \{0.106,...,1.944\}$ and $D_X$ is a cosmological distance scale depending on the data set, according to Tab.~\ref{tab:baodata}. The drag epoch $z_d$ is the time where the baryons were released from the Compton drag of the photons and hence the acoustic oscillations frozen in. From the equations in Ref.~\cite{Eisenstein:1997ik}, we calculate $z_d \simeq 1060$. 

Taking the ratio of the sound horizons cancels the dependence of the early-time cosmology before $z_*$ and thus $\Pi_i$ depends only on the expansion history between photon decoupling and today \cite{Sollerman:2009yu}. The ratio of the sound horizons at $z_d \simeq 1060$ and $z_* \simeq 1090$ depends on the expansion between these two redshifts. Since they are relatively close, we can use the best-fit values of the sound horizons from the Planck 2018 data release (TT+lowE): $r_{s}(z_*) = (144.46 \pm 0.48 )\, \mathrm{Mpc}$ and $r_{s}(z_d) = (147.21 \pm 0.48) \, \mathrm{Mpc}$ \cite{Aghanim:2018eyx}, so that,
\begin{equation}
\label{eq:rsRat}
r_{s}(z_d) / r_{s}(z_*) = 1.019 \pm 0.005.
\end{equation}
We calculate the right-hand side of \eqref{eq:CMBBAO}, $\Pi_i^\mathrm{model}$, for each bimetric model and the value of the left-hand side, $\Pi_i^\mathrm{obs}$, is obtained through the observed quantities \eqref{eq:ShiftParams} and \eqref{eq:rsRat} together with the Tab.~\ref{tab:baodata}. The $\chi^2$ for each model is calculated as,
\begin{equation}
\label{eq:chisq}
\chi_\mathrm{CMB/BAO}^2 = \sum_{ij} (\Pi^\mathrm{model}_i - \Pi^\mathrm{obs}_i) C_{ij}^{-1} (\Pi^\mathrm{model}_j - \Pi^\mathrm{obs}_j).
\end{equation}\\

\paragraph{SNIa data.} Since the properly calibrated luminosity of supernovae of type Ia is believed to be independent of redshift, they can used as standard candles, allowing us to calibrate cosmological distances. Here, we make use of binned data of the peak apparent magnitude $m_B$ from the Pantheon data set; 40 bins ranging from $z=0.014$ to $z=1.61$ \cite{Scolnic:2017caz}.\footnote{Available at \href{https://github.com/dscolnic/Pantheon/tree/master/Binned_data}{https://github.com/dscolnic/Pantheon/tree/master/Binned\_data}, last checked 2020-11-05.} $m_B$ is given by,
\begin{equation}
m_B = \mathcal{M} + 5 \log_{10} d_L,
\end{equation}
where $d_L$ is the luminosity distance rescaled by a factor of $H_0/c$ (here, $c$ is the speed of light), 
\begin{equation}
d_L = (1+z) \int_{0}^{z} \frac{dz'}{E(z')},
\end{equation}
and $\mathcal{M}$ is,
\begin{equation}
\mathcal{M} = 25 + M_B + 5 \log_{10} (c/H_0),
\end{equation}
with $M_B$ being the absolute magnitude of the supernova. Here, we marginalize over $\mathcal{M}$, see for example \cite{Goliath:2001af}, with the result,
\begin{equation}
\label{eq:chisqSNIa}
\chi^2_\mathrm{SNIa} = \widehat{\chi}^2 - B^2 /C + \ln (C / 2\pi),
\end{equation}
where,
\begin{alignat}{2}
\Delta_i &\equiv 5 \log_{10} d_L(z_i) - m_B(z_i),& \quad \widehat{\chi}^2 &\equiv \sum_{ij} \Delta_i C_{ij}^{-1} \Delta_j,\\
B &\equiv \sum_{ij} u_i C_{ij}^{-1} \Delta_j,& \quad C &\equiv \sum_{ij} u_i C_{ij}^{-1} u_j,
\end{alignat}
and
\begin{equation}
u_i = 1, \quad i \in \{1,2,...,40\}.
\end{equation}
$C_{ij}$ is the covariance matrix, including both statistical and systematic errors.

\section{Two-parameter models}
\label{sec:TwoParam}

\begin{figure}[t]
	\centering
	\includegraphics[width=\linewidth]{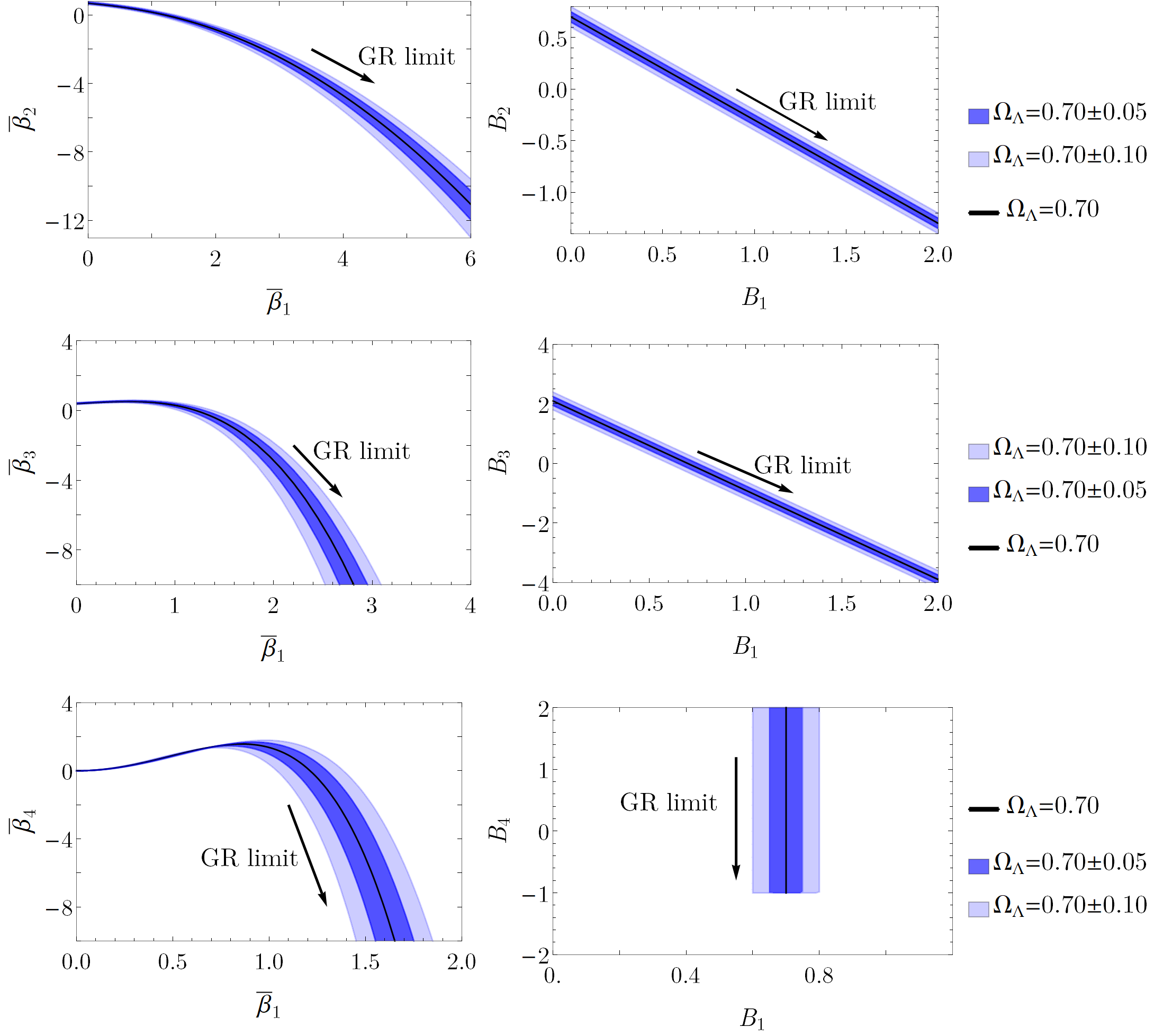}
	\caption{Viable regions in the parameter spaces of self-accelerating two-parameter models: $B_1 B_2$ (top), $B_1 B_3$ (middle), and $B_1 B_4$ (bottom). The black curves are traced out by letting $\theta \to 0$ with fixed $\omegaleff = 0.70$. Right panel: results in the $B$-parameter plane, with $B_n$ defined as in \eqref{eq:Bdef}. Left panel: the $\bar{\beta}$-parameters can be identified with the $B$-parameters of \cite{Akrami:2012vf} which is the same as the $\beta$-parameters of \cite{Konnig:2013gxa}. Indeed, the analytical prediction is consistent with \cite{Akrami:2012vf,Konnig:2013gxa} where the bimetric models are fitted to cosmological data.}
	\label{fig:specialmodels}
\end{figure}
Here, all $B$-parameters but two are set to zero. To have a real-valued cosmology in the early universe, $B_1>0$ \cite{PhysParamTh}. These models are not compatible with a working screening mechanism which require also non-vanishing $B_2$ and $B_3$. However, we include them here to connect to earlier works. As an example, in a $B_1 B_2$ model (i.e., $B_0 = B_3 = B_4 =0$), we can let $\theta$ and $\omegaleff$ be our independent parameters. Then $\alpha$, $\beta$, and $\mfphat$ can be expressed in terms of these as,
\begin{subequations}
	\begin{alignat}{2}
	\label{eq:mFPB12}
	\mfphat^2 &= \frac{1}{2} (\csc^2 \theta + 3 \sec^2 \theta) \omegaleff,& \quad B_0 &= B_3 = B_4 =0,\\
	\alpha &= \frac{1-3\tan^2 \theta}{1+3\tan^2 \theta},& \quad B_0 &= B_3 = B_4 =0,\\
	\beta &= 0,& \quad B_0 &= B_3 = B_4 =0.
	\end{alignat}
\end{subequations}
The $B$-parameters read,
\begin{subequations}
	\begin{alignat}{2}
	B_1 &= \frac{3}{2} (1 - \tan^2 \theta) \omegaleff,& \quad B_0 &= B_3 = B_4 =0,\\
	B_2 &= - \frac{1}{2} (1 - 3 \tan^2 \theta) \omegaleff,& \quad B_0 &= B_3 = B_4 =0.
	\end{alignat}
\end{subequations}
For more examples, see \cite{Luben:2020xll}. A similar exercise can be carried out for the other self-accelerating two-parameter models $B_1 B_3$ and $B_1 B_4$.

Assuming that the best fit to these models is close to a $\Lambda$CDM model, we can predict analytically the observationally viable regions in the parameter space, without making any actual data fitting. Setting $\omegaleff = 0.7$, the GR limit $\theta \to 0$ defines a curve in the parameter space along which we approach $\Lambda$CDM cosmology with $\Omega_\Lambda \simeq 0.7$. Hence, we expect our best-fit models to lie in the neighborhood of this line, see Fig.~\ref{fig:specialmodels}.

\section{Numerical details}
\subsection{Numerical errors}
Numerical errors are controlled by estimating upper limits in the derived quantities. Ultimately, we are interested in the $\chi^2$ at each point of the parameter space, which we calculate using \eqref{eq:chisq} and \eqref{eq:chisqSNIa}, depending on $E(z)$ and $I(z) \equiv \int_{0}^{z} dz'/E(z')$. We estimate an upper limit on the numerical errors of these. We introduce a discrete logarithmic grid (plus the point $z=0$) containing 200 points in the relevant redshift range $0 \leq z \leq z_*$.

We start by evaluating the equations \eqref{eq:yPoly} and \eqref{eq:BRFriedm} at $z=0$ (i.e., today) and solve for $y_0$ and $\omegamnot$. Rewriting the equations,
\begin{subequations}
	\label{eq:NormEq}
	\begin{align}
		\omegamnot &= 1 - \omegaleff + \mfphat^2 \sin^2 \theta  (1-y_0) \left[1 + \alpha(1-y_0) + \frac{\beta}{3}(1-y_0)^2\right],\\
		\label{eq:y0eq}
		0 &= -1 + \omegaleff y_0^2 + \frac{1}{3} \cos^2 \theta \, \mfphat \frac{1}{y_0} \left[1 + 2\alpha + \beta -3(\alpha+\beta)y_0 + 3 \beta y_0^2 - (1-\alpha+\beta)y_0^3\right].
	\end{align}
\end{subequations}
From $\omegamnot$, we know $\Omega_m = \omegamnot (1+z)^{3(1+w_m)}$ as a function of redshift and $y$ can be solved at each redshift from eq. \eqref{eq:yPoly}. Finally, $E$ is computed by $E(z)= \sqrt{\Omega_m(z) + \omegade(y(z))}$ \eqref{eq:BRFriedm} at each $z$.

\begin{figure}[t]
	\centering
	\includegraphics[width=0.5\linewidth]{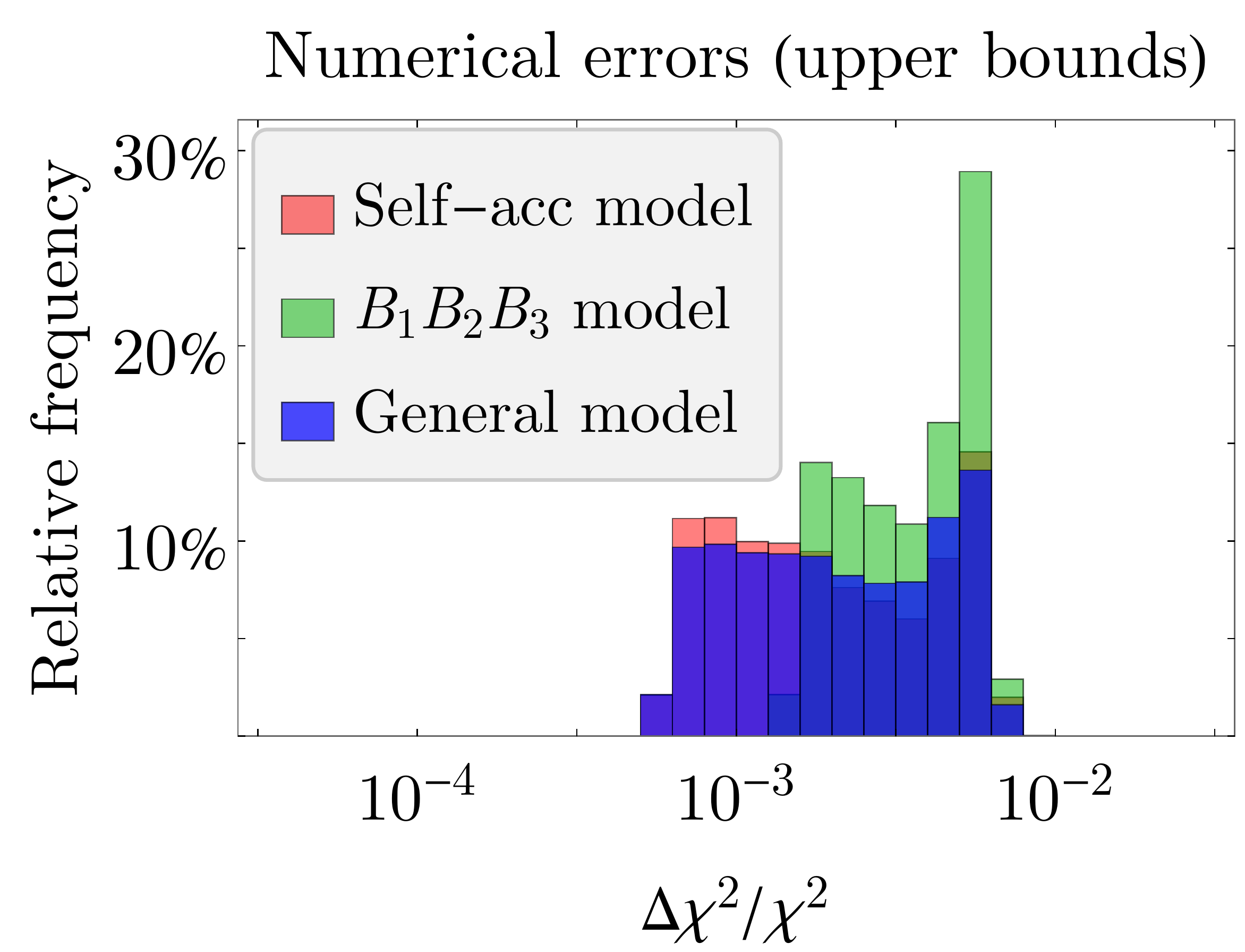}
	\caption{The distribution of relative errors in $\chi^2$ (note that these errors are upper limits). In all cases, $\Delta \chi^2 / \chi^2 < 10^{-2}$.}
	\label{fig:errfig}
\end{figure}

When solving the equations for $y_0$, $\omegamnot$, and $y$, we introduce numerical errors. We solve the equations to 15 digit precision, meaning that the relative errors in these quantities are $10^{-15}$. These numerical errors propagate to $E$ via,
\begin{subequations}
	\label{eq:DeltaE}
	\begin{alignat}{2}
		&\left[\frac{\Delta E}{E}\right]_{\mathrm{numerical \;} y}& &= \frac{1}{2} \frac{y}{\omegadef} \frac{\partial \omegadef}{\partial y} \frac{\Delta y}{y},\\
		&\left[\frac{\Delta E}{E}\right]_{\mathrm{numerical} \; \omegamnot}& &= \frac{1}{2} \frac{\Omega_m}{\omegadef} \frac{\partial \omegadef}{\partial y} \left(\frac{\partial \Omega_m}{\partial y}\right)^{-1} \frac{\Delta \omegamnot}{\omegamnot},\\
		&\left[\frac{\Delta E}{E}\right]_{\mathrm{numerical} \; y_0}& &= \frac{1}{2} \frac{\Omega_m}{\omegamnot} \frac{y_0}{\omegadef} \frac{\partial \omegadef}{\partial y} \left(\frac{\partial \Omega_m}{\partial y}\right)^{-1} \frac{\partial \omegamnot}{\partial y_0} \frac{\Delta y_0}{y_0}.		
	\end{alignat}
\end{subequations}
Here, $\Delta$ denotes the numerical error in a calculated quantity. We are now ready to show how this error affect the $\chi^2$ value. Starting with CMB/BAO, the numerical error in $\chi^2$ comes from $\Pi_i^\mathrm{model}$ which has contributions from $\Delta I_*$, $\Delta I(z_i)$, and $\Delta E$ (see eq. \eqref{eq:chisq}). Using the chain rule and the triangle inequality,
\begin{equation}
\label{eq:errPi}
	\left|\frac{\Delta \Pi_i}{\Pi_i}\right| \leq \left|\frac{\Delta I_*}{I_*}\right| + \frac{2}{3} \left|\frac{\Delta I(z_i)}{I(z_i)}\right| + \frac{1}{3} \left|\frac{\Delta E}{E}\right|,
\end{equation}
where the last term is given in \eqref{eq:DeltaE}. In the integral terms, there are two types of errors contributing: the numerical error in $E$ and the finite redshift grid size. Concerning the first type, they tend to cancel when integrating. However, as an upper limit we estimate it to be of the same magnitude as $\Delta E/E$. The errors due to the finite redshift grid size is estimated by comparing the value of the integral using a grid with 100 points and a grid with 200 points. The difference in their value compared to the value of the integral gives an upper limit on the numerical error,
\begin{equation}
	\frac{\Delta I}{I} < \frac{I_\mathrm{100 \; points}-I_\mathrm{200 \; points}}{I_\mathrm{200 \; points}}.
\end{equation}
Using the variance formula, we finally get the upper limit on the numerical error for $\chi^2$,
\begin{equation}
	\left|\frac{\Delta \chi^2_\mathrm{CMB/BAO}}{\chi^2_\mathrm{CMB/BAO}}\right| < \frac{2}{\chi^2_\mathrm{CMB/BAO}}
	\mathrm{Max} \left|\frac{\Delta \Pi_i}{\Pi_i}\right| \sum_{i=1}^{4} \left|\Pi_i \frac{\Pi_i - \Pi_\mathrm{obs,i}}{\sigma_i^2}\right|
\end{equation}
Similarly, for the SNIa calculations,
\begin{equation}
	\left|\frac{\Delta \chi^2_\mathrm{SNIa}}{\chi^2_\mathrm{SNIa}}\right| < \frac{10}{\chi^2_\mathrm{SNIa} \ln 10} \sum_{i=1}^{44} \frac{1}{\sigma_i^2} \left|\Delta_i - \frac{B}{C}\right| \left|\frac{\Delta I(z_i)}{I(z_i)}\right|,
\end{equation}
where $i=1,...,40$ and $z_1,...,z_{40}$ are the redshift of the different bins. The resulting upper limits on the numerical errors are presented in Fig.~\ref{fig:errfig}.

\subsection{Scanning details}
In the scanning process, we implement the constraints of Section~\ref{sec:BR} by assigning zero likelihood to the points in parameter space which violate these constraints. A couple of numerical problems can occur at each point in the parameter space. First, solving the equations for $y_0$ and $y$, the numerical algorithm may not find the finite branch root, that is within the range $0 < y <1$. Second, the algorithm may not be able to solve the equations to the required precision (15 digits). Both these problems can in principle be solved by refining the numerical algorithm. If any of these errors occur, we remove that point from the grid. To ensure that the final error in $\chi^2$ is small enough, we remove points with $\Delta \chi^2 / \chi^2 > 10^{-2}$.

Scanning the general models (Fig.~\ref{fig:cosmotestsGen}), we use a grid of size $30^5 = 24.3 \times 10^6$ points of which $\simeq 0.1 \, \%$ of the points were removed due to numerical problems. Scanning the self-accelerating models (Fig.~\ref{fig:cosmotestsSA}), the grid size is $41^4 \simeq 2.8 \times 10^6$ points whereof $0.6 \, \%$ were removed. The corresponding numbers for the $B_1 B_2 B_3$ models are a grid size of $80^3 \simeq 0.5 \times 10^6$ points whereof none were removed.

\bibliographystyle{JHEP}
\bibliography{biblio}

\providecommand{\href}[2]{#2}\begingroup\raggedright\begin{thebibliography}{10}

\bibitem{Volkov:2011an}
M.~S. Volkov, \emph{{Cosmological solutions with massive gravitons in the
  bigravity theory}},
  \href{http://dx.doi.org/10.1007/JHEP01(2012)035}{\emph{JHEP} {\bf 01} (2012)
  035}, [\href{http://arxiv.org/abs/1110.6153}{{\tt 1110.6153}}].

\bibitem{vonStrauss:2011mq}
M.~von Strauss, A.~Schmidt-May, J.~Enander, E.~Mortsell and S.~F. Hassan,
  \emph{{Cosmological Solutions in Bimetric Gravity and their Observational
  Tests}}, \href{http://dx.doi.org/10.1088/1475-7516/2012/03/042}{\emph{JCAP}
  {\bf 1203} (2012) 042}, [\href{http://arxiv.org/abs/1111.1655}{{\tt
  1111.1655}}].

\bibitem{Comelli:2011zm}
D.~Comelli, M.~Crisostomi, F.~Nesti and L.~Pilo, \emph{{FRW Cosmology in Ghost
  Free Massive Gravity}}, \href{http://dx.doi.org/10.1007/JHEP06(2012)020,
  10.1007/JHEP03(2012)067}{\emph{JHEP} {\bf 03} (2012) 067},
  [\href{http://arxiv.org/abs/1111.1983}{{\tt 1111.1983}}].

\bibitem{Volkov:2012wp}
M.~S. Volkov, \emph{{Hairy black holes in the ghost-free bigravity theory}},
  \href{http://dx.doi.org/10.1103/PhysRevD.85.124043}{\emph{Phys. Rev.} {\bf
  D85} (2012) 124043}, [\href{http://arxiv.org/abs/1202.6682}{{\tt
  1202.6682}}].

\bibitem{Volkov:2012zb}
M.~S. Volkov, \emph{{Exact self-accelerating cosmologies in the ghost-free
  massive gravity -- the detailed derivation}},
  \href{http://dx.doi.org/10.1103/PhysRevD.86.104022}{\emph{Phys. Rev. D} {\bf
  86} (2012) 104022}, [\href{http://arxiv.org/abs/1207.3723}{{\tt 1207.3723}}].

\bibitem{Akrami:2012vf}
Y.~Akrami, T.~S. Koivisto and M.~Sandstad, \emph{{Accelerated expansion from
  ghost-free bigravity: a statistical analysis with improved generality}},
  \href{http://dx.doi.org/10.1007/JHEP03(2013)099}{\emph{JHEP} {\bf 03} (2013)
  099}, [\href{http://arxiv.org/abs/1209.0457}{{\tt 1209.0457}}].

\bibitem{Volkov:2013roa}
M.~S. Volkov, \emph{{Self-accelerating cosmologies and hairy black holes in
  ghost-free bigravity and massive gravity}},
  \href{http://dx.doi.org/10.1088/0264-9381/30/18/184009}{\emph{Class. Quant.
  Grav.} {\bf 30} (2013) 184009}, [\href{http://arxiv.org/abs/1304.0238}{{\tt
  1304.0238}}].

\bibitem{Konnig:2013gxa}
F.~Koennig, A.~Patil and L.~Amendola, \emph{{Viable cosmological solutions in
  massive bimetric gravity}},
  \href{http://dx.doi.org/10.1088/1475-7516/2014/03/029}{\emph{JCAP} {\bf 1403}
  (2014) 029}, [\href{http://arxiv.org/abs/1312.3208}{{\tt 1312.3208}}].

\bibitem{tHooft:1979rat}
G.~'t~Hooft, \emph{{Naturalness, chiral symmetry, and spontaneous chiral
  symmetry breaking}},
  \href{http://dx.doi.org/10.1007/978-1-4684-7571-5_9}{\emph{NATO Sci. Ser. B}
  {\bf 59} (1980) 135--157}.

\bibitem{Mortsell:2018mfj}
E.~Mörtsell and S.~Dhawan, \emph{{Does the Hubble constant tension call for
  new physics?}},
  \href{http://dx.doi.org/10.1088/1475-7516/2018/09/025}{\emph{JCAP} {\bf 1809}
  (2018) 025}, [\href{http://arxiv.org/abs/1801.07260}{{\tt 1801.07260}}].

\bibitem{Babichev:2013pfa}
E.~Babichev and M.~Crisostomi, \emph{{Restoring general relativity in massive
  bigravity theory}},
  \href{http://dx.doi.org/10.1103/PhysRevD.88.084002}{\emph{Phys. Rev.} {\bf
  D88} (2013) 084002}, [\href{http://arxiv.org/abs/1307.3640}{{\tt
  1307.3640}}].

\bibitem{Enander:2015kda}
J.~Enander and E.~Mortsell, \emph{{On stars, galaxies and black holes in
  massive bigravity}},
  \href{http://dx.doi.org/10.1088/1475-7516/2015/11/023}{\emph{JCAP} {\bf 1511}
  (2015) 023}, [\href{http://arxiv.org/abs/1507.00912}{{\tt 1507.00912}}].

\bibitem{Platscher:2018voh}
M.~Platscher, J.~Smirnov, S.~Meyer and M.~Bartelmann, \emph{{Long Range Effects
  in Gravity Theories with Vainshtein Screening}},
  \href{http://dx.doi.org/10.1088/1475-7516/2018/12/009}{\emph{JCAP} {\bf 12}
  (2018) 009}, [\href{http://arxiv.org/abs/1809.05318}{{\tt 1809.05318}}].

\bibitem{Comelli:2012db}
D.~Comelli, M.~Crisostomi and L.~Pilo, \emph{{Perturbations in Massive Gravity
  Cosmology}}, \href{http://dx.doi.org/10.1007/JHEP06(2012)085}{\emph{JHEP}
  {\bf 06} (2012) 085}, [\href{http://arxiv.org/abs/1202.1986}{{\tt
  1202.1986}}].

\bibitem{Khosravi:2012rk}
N.~Khosravi, H.~R. Sepangi and S.~Shahidi, \emph{{Massive cosmological scalar
  perturbations}},
  \href{http://dx.doi.org/10.1103/PhysRevD.86.043517}{\emph{Phys. Rev.} {\bf
  D86} (2012) 043517}, [\href{http://arxiv.org/abs/1202.2767}{{\tt
  1202.2767}}].

\bibitem{Berg:2012kn}
M.~Berg, I.~Buchberger, J.~Enander, E.~Mortsell and S.~Sjors, \emph{{Growth
  Histories in Bimetric Massive Gravity}},
  \href{http://dx.doi.org/10.1088/1475-7516/2012/12/021}{\emph{JCAP} {\bf 1212}
  (2012) 021}, [\href{http://arxiv.org/abs/1206.3496}{{\tt 1206.3496}}].

\bibitem{Sakakihara:2012iq}
Y.~Sakakihara, J.~Soda and T.~Takahashi, \emph{{On Cosmic No-hair in Bimetric
  Gravity and the Higuchi Bound}},
  \href{http://dx.doi.org/10.1093/ptep/ptt004}{\emph{PTEP} {\bf 2013} (2013)
  033E02}, [\href{http://arxiv.org/abs/1211.5976}{{\tt 1211.5976}}].

\bibitem{Konnig:2014dna}
F.~Könnig and L.~Amendola, \emph{{Instability in a minimal bimetric gravity
  model}}, \href{http://dx.doi.org/10.1103/PhysRevD.90.044030}{\emph{Phys.
  Rev.} {\bf D90} (2014) 044030}, [\href{http://arxiv.org/abs/1402.1988}{{\tt
  1402.1988}}].

\bibitem{Comelli:2014bqa}
D.~Comelli, M.~Crisostomi and L.~Pilo, \emph{{FRW Cosmological Perturbations in
  Massive Bigravity}},
  \href{http://dx.doi.org/10.1103/PhysRevD.90.084003}{\emph{Phys. Rev.} {\bf
  D90} (2014) 084003}, [\href{http://arxiv.org/abs/1403.5679}{{\tt
  1403.5679}}].

\bibitem{DeFelice:2014nja}
A.~De~Felice, A.~E. Gümrükçüoğlu, S.~Mukohyama, N.~Tanahashi and
  T.~Tanaka, \emph{{Viable cosmology in bimetric theory}},
  \href{http://dx.doi.org/10.1088/1475-7516/2014/06/037}{\emph{JCAP} {\bf 1406}
  (2014) 037}, [\href{http://arxiv.org/abs/1404.0008}{{\tt 1404.0008}}].

\bibitem{Solomon:2014dua}
A.~R. Solomon, Y.~Akrami and T.~S. Koivisto, \emph{{Linear growth of structure
  in massive bigravity}},
  \href{http://dx.doi.org/10.1088/1475-7516/2014/10/066}{\emph{JCAP} {\bf 1410}
  (2014) 066}, [\href{http://arxiv.org/abs/1404.4061}{{\tt 1404.4061}}].

\bibitem{Konnig:2014xva}
F.~Koennig, Y.~Akrami, L.~Amendola, M.~Motta and A.~R. Solomon, \emph{{Stable
  and unstable cosmological models in bimetric massive gravity}},
  \href{http://dx.doi.org/10.1103/PhysRevD.90.124014}{\emph{Phys. Rev.} {\bf
  D90} (2014) 124014}, [\href{http://arxiv.org/abs/1407.4331}{{\tt
  1407.4331}}].

\bibitem{Lagos:2014lca}
M.~Lagos and P.~G. Ferreira, \emph{{Cosmological perturbations in massive
  bigravity}},
  \href{http://dx.doi.org/10.1088/1475-7516/2014/12/026}{\emph{JCAP} {\bf 1412}
  (2014) 026}, [\href{http://arxiv.org/abs/1410.0207}{{\tt 1410.0207}}].

\bibitem{Konnig:2015lfa}
F.~Könnig, \emph{{Higuchi Ghosts and Gradient Instabilities in Bimetric
  Gravity}}, \href{http://dx.doi.org/10.1103/PhysRevD.91.104019}{\emph{Phys.
  Rev.} {\bf D91} (2015) 104019}, [\href{http://arxiv.org/abs/1503.07436}{{\tt
  1503.07436}}].

\bibitem{Aoki:2015xqa}
K.~Aoki, K.-i. Maeda and R.~Namba, \emph{{Stability of the Early Universe in
  Bigravity Theory}},
  \href{http://dx.doi.org/10.1103/PhysRevD.92.044054}{\emph{Phys. Rev.} {\bf
  D92} (2015) 044054}, [\href{http://arxiv.org/abs/1506.04543}{{\tt
  1506.04543}}].

\bibitem{Mortsell:2015exa}
E.~Mortsell and J.~Enander, \emph{{Scalar instabilities in bimetric gravity:
  The Vainshtein mechanism and structure formation}},
  \href{http://dx.doi.org/10.1088/1475-7516/2015/10/044}{\emph{JCAP} {\bf 1510}
  (2015) 044}, [\href{http://arxiv.org/abs/1506.04977}{{\tt 1506.04977}}].

\bibitem{Akrami:2015qga}
Y.~Akrami, S.~F. Hassan, F.~Könnig, A.~Schmidt-May and A.~R. Solomon,
  \emph{{Bimetric gravity is cosmologically viable}},
  \href{http://dx.doi.org/10.1016/j.physletb.2015.06.062}{\emph{Phys. Lett.}
  {\bf B748} (2015) 37--44}, [\href{http://arxiv.org/abs/1503.07521}{{\tt
  1503.07521}}].

\bibitem{Hogas:2019ywm}
M.~Högås, F.~Torsello and E.~Mörtsell, \emph{{On the stability of bimetric
  structure formation}},
  \href{http://dx.doi.org/10.1088/1475-7516/2020/04/046}{\emph{JCAP} {\bf 04}
  (2020) 046}, [\href{http://arxiv.org/abs/1910.01651}{{\tt 1910.01651}}].

\bibitem{Luben:2019yyx}
M.~Lüben, A.~Schmidt-May and J.~Smirnov, \emph{{Vainshtein Screening in
  Bimetric Cosmology}},  \href{http://arxiv.org/abs/1912.09449}{{\tt
  1912.09449}}.

\bibitem{Kocic:2018ddp}
M.~Kocic, \emph{{Geometric mean of bimetric spacetimes}},
  \href{http://dx.doi.org/10.1088/1361-6382/abdf28}{\emph{Class. Quant. Grav.}
  {\bf 38} (2021) 075023}, [\href{http://arxiv.org/abs/1803.09752}{{\tt
  1803.09752}}].

\bibitem{Kocic:2018yvr}
M.~Kocic, \emph{{Causal propagation of constraints in bimetric relativity in
  standard 3+1 form}},
  \href{http://dx.doi.org/10.1007/JHEP10(2019)219}{\emph{JHEP} {\bf 10} (2019)
  219}, [\href{http://arxiv.org/abs/1804.03659}{{\tt 1804.03659}}].

\bibitem{Kocic:2019zdy}
M.~Kocic, A.~Lundkvist and F.~Torsello, \emph{{On the ratio of lapses in
  bimetric relativity}},
  \href{http://dx.doi.org/10.1088/1361-6382/ab497a}{\emph{Class. Quant. Grav.}
  {\bf 36} (2019) 225013}, [\href{http://arxiv.org/abs/1903.09646}{{\tt
  1903.09646}}].

\bibitem{Torsello:2019tgc}
F.~Torsello, M.~Kocic, M.~Högås and E.~Mörtsell, \emph{{Covariant BSSN
  formulation in bimetric relativity}},
  \href{http://dx.doi.org/10.1088/1361-6382/ab56fc}{\emph{Class. Quant. Grav.}
  {\bf 37} (2020) 025013}, [\href{http://arxiv.org/abs/1904.07869}{{\tt
  1904.07869}}].

\bibitem{Kocic:2019gxl}
M.~Kocic, F.~Torsello, M.~Högås and E.~Mortsell, \emph{{Spherical dust
  collapse in bimetric relativity: Bimetric polytropes}},
  \href{http://arxiv.org/abs/1904.08617}{{\tt 1904.08617}}.

\bibitem{Torsello:2019jdg}
F.~Torsello, \emph{{The mean gauges in bimetric relativity}},
  \href{http://dx.doi.org/10.1088/1361-6382/ab4ccf}{\emph{Class. Quant. Grav.}
  {\bf 36} (2019) 235010}, [\href{http://arxiv.org/abs/1904.09297}{{\tt
  1904.09297}}].

\bibitem{Torsello:2019wyp}
F.~Torsello, \emph{{$\mathtt{bimEX}$: A Mathematica package for exact
  computations in $3+1$ bimetric relativity}},
  \href{http://dx.doi.org/10.1016/j.cpc.2019.106948}{\emph{Comput. Phys.
  Commun.} {\bf 247} (2020) 106948},
  [\href{http://arxiv.org/abs/1904.10464}{{\tt 1904.10464}}].

\bibitem{Kocic:2020pnm}
M.~Kocic, F.~Torsello, M.~Högås and E.~Mörtsell, \emph{{Initial data and
  first evolutions of dust clouds in bimetric relativity}},
  \href{http://dx.doi.org/10.1088/1361-6382/ab87d8}{\emph{Class. Quant. Grav.}
  {\bf 37} (2020) 165010}.

\bibitem{PhysParamTh}
M.~H\"og\r{a}s and E.~M\"ortsell, \emph{{Analytical constraints on bimetric
  gravity}}, \href{http://dx.doi.org/10.1088/1475-7516/2021/05/001}{\emph{JCAP}
  {\bf 05} (2021) 001}, [\href{http://arxiv.org/abs/2101.08794}{{\tt
  2101.08794}}].

\bibitem{Luben:2020xll}
M.~Lüben, A.~Schmidt-May and J.~Weller, \emph{{Physical parameter space of
  bimetric theory and SN1a constraints}},
  \href{http://arxiv.org/abs/2003.03382}{{\tt 2003.03382}}.

\bibitem{Dhawan:2017leu}
S.~Dhawan, A.~Goobar, E.~Mörtsell, R.~Amanullah and U.~Feindt,
  \emph{{Narrowing down the possible explanations of cosmic acceleration with
  geometric probes}},
  \href{http://dx.doi.org/10.1088/1475-7516/2017/07/040}{\emph{JCAP} {\bf 1707}
  (2017) 040}, [\href{http://arxiv.org/abs/1705.05768}{{\tt 1705.05768}}].

\bibitem{Lindner:2020eez}
M.~Lindner, K.~Max, M.~Platscher and J.~Rezacek, \emph{{Probing alternative
  cosmologies through the inverse distance ladder}},
  \href{http://arxiv.org/abs/2002.01487}{{\tt 2002.01487}}.

\bibitem{DeFelice:2013nba}
A.~De~Felice, T.~Nakamura and T.~Tanaka, \emph{{Possible existence of viable
  models of bi-gravity with detectable graviton oscillations by gravitational
  wave detectors}}, \href{http://dx.doi.org/10.1093/ptep/ptu024}{\emph{PTEP}
  {\bf 2014} (2014) 043E01}, [\href{http://arxiv.org/abs/1304.3920}{{\tt
  1304.3920}}].

\bibitem{Fasiello:2015csa}
M.~Fasiello and R.~H. Ribeiro, \emph{{Mild bounds on bigravity from primordial
  gravitational waves}},
  \href{http://dx.doi.org/10.1088/1475-7516/2015/07/027}{\emph{JCAP} {\bf 1507}
  (2015) 027}, [\href{http://arxiv.org/abs/1505.00404}{{\tt 1505.00404}}].

\bibitem{Cusin:2015pya}
G.~Cusin, R.~Durrer, P.~Guarato and M.~Motta, \emph{{Inflationary perturbations
  in bimetric gravity}},
  \href{http://dx.doi.org/10.1088/1475-7516/2015/09/043}{\emph{JCAP} {\bf 09}
  (2015) 043}, [\href{http://arxiv.org/abs/1505.01091}{{\tt 1505.01091}}].

\bibitem{Max:2017flc}
K.~Max, M.~Platscher and J.~Smirnov, \emph{{Gravitational Wave Oscillations in
  Bigravity}},
  \href{http://dx.doi.org/10.1103/PhysRevLett.119.111101}{\emph{Phys. Rev.
  Lett.} {\bf 119} (2017) 111101}, [\href{http://arxiv.org/abs/1703.07785}{{\tt
  1703.07785}}].

\bibitem{Luben:2018ekw}
M.~Lüben, E.~Mörtsell and A.~Schmidt-May, \emph{{Bimetric cosmology is
  compatible with local tests of gravity}},
  \href{http://arxiv.org/abs/1812.08686}{{\tt 1812.08686}}.

\bibitem{Sjors:2011iv}
S.~Sjors and E.~Mortsell, \emph{{Spherically Symmetric Solutions in Massive
  Gravity and Constraints from Galaxies}},
  \href{http://dx.doi.org/10.1007/JHEP02(2013)080}{\emph{JHEP} {\bf 02} (2013)
  080}, [\href{http://arxiv.org/abs/1111.5961}{{\tt 1111.5961}}].

\bibitem{Enander:2013kza}
J.~Enander and E.~Mörtsell, \emph{{Strong lensing constraints on bimetric
  massive gravity}},
  \href{http://dx.doi.org/10.1007/JHEP10(2013)031}{\emph{JHEP} {\bf 10} (2013)
  031}, [\href{http://arxiv.org/abs/1306.1086}{{\tt 1306.1086}}].

\bibitem{Boulware:1973my}
D.~G. Boulware and S.~Deser, \emph{{Can gravitation have a finite range?}},
  \href{http://dx.doi.org/10.1103/PhysRevD.6.3368}{\emph{Phys. Rev.} {\bf D6}
  (1972) 3368--3382}.

\bibitem{Hassan:2011zd}
S.~F. Hassan and R.~A. Rosen, \emph{{Bimetric Gravity from Ghost-free Massive
  Gravity}}, \href{http://dx.doi.org/10.1007/JHEP02(2012)126}{\emph{JHEP} {\bf
  02} (2012) 126}, [\href{http://arxiv.org/abs/1109.3515}{{\tt 1109.3515}}].

\bibitem{deRham:2010ik}
C.~de~Rham and G.~Gabadadze, \emph{{Generalization of the Fierz-Pauli Action}},
  \href{http://dx.doi.org/10.1103/PhysRevD.82.044020}{\emph{Phys. Rev.} {\bf
  D82} (2010) 044020}, [\href{http://arxiv.org/abs/1007.0443}{{\tt
  1007.0443}}].

\bibitem{deRham:2010kj}
C.~de~Rham, G.~Gabadadze and A.~J. Tolley, \emph{{Resummation of Massive
  Gravity}},
  \href{http://dx.doi.org/10.1103/PhysRevLett.106.231101}{\emph{Phys. Rev.
  Lett.} {\bf 106} (2011) 231101}, [\href{http://arxiv.org/abs/1011.1232}{{\tt
  1011.1232}}].

\bibitem{Hassan:2012wr}
S.~F. Hassan, A.~Schmidt-May and M.~von Strauss, \emph{{On Consistent Theories
  of Massive Spin-2 Fields Coupled to Gravity}},
  \href{http://dx.doi.org/10.1007/JHEP05(2013)086}{\emph{JHEP} {\bf 05} (2013)
  086}, [\href{http://arxiv.org/abs/1208.1515}{{\tt 1208.1515}}].

\bibitem{Hassan:2014vja}
S.~F. Hassan, A.~Schmidt-May and M.~von Strauss, \emph{{Particular Solutions in
  Bimetric Theory and Their Implications}},
  \href{http://dx.doi.org/10.1142/S0218271814430020}{\emph{Int. J. Mod. Phys.}
  {\bf D23} (2014) 1443002}, [\href{http://arxiv.org/abs/1407.2772}{{\tt
  1407.2772}}].

\bibitem{Vainshtein:1972sx}
A.~I. Vainshtein, \emph{{To the problem of nonvanishing gravitation mass}},
  \href{http://dx.doi.org/10.1016/0370-2693(72)90147-5}{\emph{Phys. Lett. B}
  {\bf 39} (1972) 393--394}.

\bibitem{Higuchi:1986py}
A.~Higuchi, \emph{{Forbidden Mass Range for Spin-2 Field Theory in De Sitter
  Space-time}},
  \href{http://dx.doi.org/10.1016/0550-3213(87)90691-2}{\emph{Nucl. Phys. B}
  {\bf 282} (1987) 397--436}.

\bibitem{Will:2014kxa}
C.~M. Will, \emph{{The Confrontation between General Relativity and
  Experiment}}, \href{http://dx.doi.org/10.12942/lrr-2014-4}{\emph{Living Rev.
  Rel.} {\bf 17} (2014) 4}, [\href{http://arxiv.org/abs/1403.7377}{{\tt
  1403.7377}}].

\bibitem{Schwab:2009nz}
J.~Schwab, A.~S. Bolton and S.~A. Rappaport, \emph{{Galaxy-Scale Strong Lensing
  Tests of Gravity and Geometric Cosmology: Constraints and Systematic
  Limitations}},
  \href{http://dx.doi.org/10.1088/0004-637X/708/1/750}{\emph{Astrophys. J.}
  {\bf 708} (2010) 750--757}, [\href{http://arxiv.org/abs/0907.4992}{{\tt
  0907.4992}}].

\bibitem{Beutler_2011}
F.~Beutler, C.~Blake, M.~Colless, D.~H. Jones, L.~Staveley-Smith, L.~Campbell
  et~al., \emph{The 6df galaxy survey: baryon acoustic oscillations and the
  local hubble constant},
  \href{http://dx.doi.org/10.1111/j.1365-2966.2011.19250.x}{\emph{Monthly
  Notices of the Royal Astronomical Society} {\bf 416} (Jul, 2011)
  3017–3032}.

\bibitem{Ross:2014qpa}
A.~J. Ross, L.~Samushia, C.~Howlett, W.~J. Percival, A.~Burden and M.~Manera,
  \emph{{The clustering of the SDSS DR7 main Galaxy sample \textendash I. A
  4~per cent distance measure at $z~=~0.15$}},
  \href{http://dx.doi.org/10.1093/mnras/stv154}{\emph{Mon. Not. Roy. Astron.
  Soc.} {\bf 449} (2015) 835--847}, [\href{http://arxiv.org/abs/1409.3242}{{\tt
  1409.3242}}].

\bibitem{Alam:2016hwk}
{\scshape BOSS} collaboration, S.~Alam et~al., \emph{{The clustering of
  galaxies in the completed SDSS-III Baryon Oscillation Spectroscopic Survey:
  cosmological analysis of the DR12 galaxy sample}},
  \href{http://dx.doi.org/10.1093/mnras/stx721}{\emph{Mon. Not. Roy. Astron.
  Soc.} {\bf 470} (2017) 2617--2652},
  [\href{http://arxiv.org/abs/1607.03155}{{\tt 1607.03155}}].

\bibitem{Bautista:2017wwp}
J.~E. Bautista et~al., \emph{{The SDSS-IV extended Baryon Oscillation
  Spectroscopic Survey: Baryon Acoustic Oscillations at redshift of 0.72 with
  the DR14 Luminous Red Galaxy Sample}},
  \href{http://dx.doi.org/10.3847/1538-4357/aacea5}{\emph{Astrophys. J.} {\bf
  863} (2018) 110}, [\href{http://arxiv.org/abs/1712.08064}{{\tt 1712.08064}}].

\bibitem{Zhao:2018gvb}
G.-B. Zhao et~al., \emph{{The clustering of the SDSS-IV extended Baryon
  Oscillation Spectroscopic Survey DR14 quasar sample: a tomographic
  measurement of cosmic structure growth and expansion rate based on optimal
  redshift weights}}, \href{http://dx.doi.org/10.1093/mnras/sty2845}{\emph{Mon.
  Not. Roy. Astron. Soc.} {\bf 482} (2019) 3497--3513},
  [\href{http://arxiv.org/abs/1801.03043}{{\tt 1801.03043}}].

\bibitem{Aghanim:2018eyx}
{\scshape Planck} collaboration, N.~Aghanim et~al., \emph{{Planck 2018 results.
  VI. Cosmological parameters}},
  \href{http://dx.doi.org/10.1051/0004-6361/201833910}{\emph{Astron.
  Astrophys.} {\bf 641} (2020) A6},
  [\href{http://arxiv.org/abs/1807.06209}{{\tt 1807.06209}}].

\bibitem{Chen:2018dbv}
L.~Chen, Q.-G. Huang and K.~Wang, \emph{{Distance Priors from Planck Final
  Release}}, \href{http://dx.doi.org/10.1088/1475-7516/2019/02/028}{\emph{JCAP}
  {\bf 02} (2019) 028}, [\href{http://arxiv.org/abs/1808.05724}{{\tt
  1808.05724}}].

\bibitem{Scolnic:2017caz}
D.~Scolnic et~al., \emph{{The Complete Light-curve Sample of Spectroscopically
  Confirmed SNe Ia from Pan-STARRS1 and Cosmological Constraints from the
  Combined Pantheon Sample}},
  \href{http://dx.doi.org/10.3847/1538-4357/aab9bb}{\emph{Astrophys. J.} {\bf
  859} (2018) 101}, [\href{http://arxiv.org/abs/1710.00845}{{\tt 1710.00845}}].

\bibitem{Allen:2004cd}
S.~Allen, R.~Schmidt, H.~Ebeling, A.~Fabian and L.~van Speybroeck,
  \emph{{Constraints on dark energy from Chandra observations of the largest
  relaxed galaxy clusters}},
  \href{http://dx.doi.org/10.1111/j.1365-2966.2004.08080.x}{\emph{Mon. Not.
  Roy. Astron. Soc.} {\bf 353} (2004) 457},
  [\href{http://arxiv.org/abs/astro-ph/0405340}{{\tt astro-ph/0405340}}].

\bibitem{Macquart2020}
J.~Macquart, J.~Prochaska, M.~McQuinn et~al., \emph{{A census of baryons in the
  Universe from localized fast radio bursts}},
  \href{http://dx.doi.org/10.1038/s41586-020-2300-2}{\emph{Nature} {\bf 581}
  (2020) 391--395}.

\bibitem{Liddle:2007fy}
A.~R. Liddle, \emph{{Information criteria for astrophysical model selection}},
  \href{http://dx.doi.org/10.1111/j.1745-3933.2007.00306.x}{\emph{Mon. Not.
  Roy. Astron. Soc.} {\bf 377} (2007) L74--L78},
  [\href{http://arxiv.org/abs/astro-ph/0701113}{{\tt astro-ph/0701113}}].

\bibitem{Ade:2015xua}
{\scshape Planck} collaboration, P.~Ade et~al., \emph{{Planck 2015 results.
  XIII. Cosmological parameters}},
  \href{http://dx.doi.org/10.1051/0004-6361/201525830}{\emph{Astron.
  Astrophys.} {\bf 594} (2016) A13},
  [\href{http://arxiv.org/abs/1502.01589}{{\tt 1502.01589}}].

\bibitem{Riess:2016jrr}
A.~G. Riess et~al., \emph{{A 2.4\% Determination of the Local Value of the
  Hubble Constant}},
  \href{http://dx.doi.org/10.3847/0004-637X/826/1/56}{\emph{Astrophys. J.} {\bf
  826} (2016) 56}, [\href{http://arxiv.org/abs/1604.01424}{{\tt 1604.01424}}].

\bibitem{Caravano:2021aum}
A.~Caravano, M.~L\"uben and J.~Weller, \emph{{Combining cosmological and local
  bounds on bimetric theory}},  \href{http://arxiv.org/abs/2101.08791}{{\tt
  2101.08791}}.

\bibitem{Ade:2015rim}
{\scshape Planck} collaboration, P.~Ade et~al., \emph{{Planck 2015 results.
  XIV. Dark energy and modified gravity}},
  \href{http://dx.doi.org/10.1051/0004-6361/201525814}{\emph{Astron.
  Astrophys.} {\bf 594} (2016) A14},
  [\href{http://arxiv.org/abs/1502.01590}{{\tt 1502.01590}}].

\bibitem{Kosowsky:2002zt}
A.~Kosowsky, M.~Milosavljevic and R.~Jimenez, \emph{{Efficient cosmological
  parameter estimation from microwave background anisotropies}},
  \href{http://dx.doi.org/10.1103/PhysRevD.66.063007}{\emph{Phys. Rev. D} {\bf
  66} (2002) 063007}, [\href{http://arxiv.org/abs/astro-ph/0206014}{{\tt
  astro-ph/0206014}}].

\bibitem{Wang:2007mza}
Y.~Wang and P.~Mukherjee, \emph{{Observational Constraints on Dark Energy and
  Cosmic Curvature}},
  \href{http://dx.doi.org/10.1103/PhysRevD.76.103533}{\emph{Phys. Rev. D} {\bf
  76} (2007) 103533}, [\href{http://arxiv.org/abs/astro-ph/0703780}{{\tt
  astro-ph/0703780}}].

\bibitem{Hu:1995en}
W.~Hu and N.~Sugiyama, \emph{{Small scale cosmological perturbations: An
  Analytic approach}}, \href{http://dx.doi.org/10.1086/177989}{\emph{Astrophys.
  J.} {\bf 471} (1996) 542--570},
  [\href{http://arxiv.org/abs/astro-ph/9510117}{{\tt astro-ph/9510117}}].

\bibitem{Efstathiou:1998xx}
G.~Efstathiou and J.~Bond, \emph{{Cosmic confusion: Degeneracies among
  cosmological parameters derived from measurements of microwave background
  anisotropies}},
  \href{http://dx.doi.org/10.1046/j.1365-8711.1999.02274.x}{\emph{Mon. Not.
  Roy. Astron. Soc.} {\bf 304} (1999) 75--97},
  [\href{http://arxiv.org/abs/astro-ph/9807103}{{\tt astro-ph/9807103}}].

\bibitem{Komatsu:2008hk}
{\scshape WMAP} collaboration, E.~Komatsu et~al., \emph{{Five-Year Wilkinson
  Microwave Anisotropy Probe (WMAP) Observations: Cosmological
  Interpretation}},
  \href{http://dx.doi.org/10.1088/0067-0049/180/2/330}{\emph{Astrophys. J.
  Suppl.} {\bf 180} (2009) 330--376},
  [\href{http://arxiv.org/abs/0803.0547}{{\tt 0803.0547}}].

\bibitem{Kunz:2007rk}
M.~Kunz, \emph{{The dark degeneracy: On the number and nature of dark
  components}}, \href{http://dx.doi.org/10.1103/PhysRevD.80.123001}{\emph{Phys.
  Rev. D} {\bf 80} (2009) 123001},
  [\href{http://arxiv.org/abs/astro-ph/0702615}{{\tt astro-ph/0702615}}].

\bibitem{Elgaroy:2007bv}
O.~Elgaroy and T.~Multamaki, \emph{{On using the CMB shift parameter in tests
  of models of dark energy}},
  \href{http://dx.doi.org/10.1051/0004-6361:20077292}{\emph{Astron. Astrophys.}
  {\bf 471} (2007) 65}, [\href{http://arxiv.org/abs/astro-ph/0702343}{{\tt
  astro-ph/0702343}}].

\bibitem{Sollerman:2009yu}
J.~Sollerman et~al., \emph{{First-Year Sloan Digital Sky Survey-II (SDSS-II)
  Supernova Results: Constraints on Non-Standard Cosmological Models}},
  \href{http://dx.doi.org/10.1088/0004-637X/703/2/1374}{\emph{Astrophys. J.}
  {\bf 703} (2009) 1374--1385}, [\href{http://arxiv.org/abs/0908.4276}{{\tt
  0908.4276}}].

\bibitem{Eisenstein:1997ik}
D.~J. Eisenstein and W.~Hu, \emph{{Baryonic features in the matter transfer
  function}}, \href{http://dx.doi.org/10.1086/305424}{\emph{Astrophys. J.} {\bf
  496} (1998) 605}, [\href{http://arxiv.org/abs/astro-ph/9709112}{{\tt
  astro-ph/9709112}}].

\bibitem{Goliath:2001af}
M.~Goliath, R.~Amanullah, P.~Astier, A.~Goobar and R.~Pain, \emph{{Supernovae
  and the nature of the dark energy}},
  \href{http://dx.doi.org/10.1051/0004-6361:20011398}{\emph{Astron. Astrophys.}
  {\bf 380} (2001) 6--18}, [\href{http://arxiv.org/abs/astro-ph/0104009}{{\tt
  astro-ph/0104009}}].

\end{thebibliography}\endgroup

\end{document}